\documentstyle [psfig]{l-aa}
\newcommand {\li} {Li\,{\sc i}\,$\lambda 6708$\AA }  
\newcommand {\etl} {\mbox{et al.}}
\newcommand {\ha} {H$\alpha$}
\begin{document}

\thesaurus{06(08.06.2, 08.12.1, 08.16.5, 13.25.5)}

\title{Search for young low-mass stars in a ROSAT selected sample
south of the Taurus-Auriga molecular clouds\thanks{Based on observations 
made with the Isaac Newton Telescope operated on the island of
La~Palma by the Royal Greenwich Observatory in the Spanish Observatorio del
Roque de los Muchachos of the Instituto de Astrof\'\i sica de Canarias
and with the ESO\,1.52m telescope on La Silla, Chile, operated by the
European Southern Observatory.} \fnmsep \thanks{Tables~1,\,2,\,3,\,4 are also available in electronic form at the CDS via anonymous ftp to cdsarc.u-strasbg.fr (130.79.128.5) or via http://cdsweb.u-strasbg.fr/Abstract.html}} 
 
\author{A.~Magazz\`u\inst{1} \and E.L.~Mart\'\i n\inst{2} \and M.F. Sterzik\inst{3} 
\and R. Neuh\"auser\inst{3} \and E. Covino\inst{4} \and J.M. Alcal\'a\inst{3}}

\offprints{A. Magazz\`u}

\institute{Osservatorio Astrofisico di Catania, Citt\`a Universitaria, I-95125 Catania, Italy 
\and Instituto de Astrof\'\i sica de Canarias, E-38200 La Laguna, Tenerife, Spain 
\and Max-Planck-Institut f\"ur Extraterrestrische Physik, D-85740 Garching, Germany 
\and Osservatorio Astronomico di Capodimonte, Via Moiariello 16, I-80131 Napoli, Italy} 

\date {Received 29 July; accepted 3 December, 1996} 

\maketitle

\markboth{A.~Magazz\`u et al.: Young low-mass stars south of Taurus-Auriga}{A.~Magazz\`u et al.: Young low-mass stars south of Taurus-Auriga}

\begin{abstract}

We present results of intermediate resolution spectroscopy of 131
optical counterparts to 115 ROSAT All-Sky Survey X-ray sources south of
the Taurus-Auriga dark cloud complex. These objects have been selected
as candidate young stars from a total of 1084 ROSAT sources in a $\sim
300$ square degree area. We identify 30 objects  as low-mass PMS stars
on the basis of the \li\ doublet in their spectrum, a signature of
their young age. All these stars have a spectral type later than F7 and
show spectral characteristics typical of weak-line and post-T~Tauri
stars. The presence of young objects several parsecs away from the
regions of ongoing star formation is discussed in the light of the
current models of T~Tauri dispersal.


\end{abstract}

\section {Introduction}

The study of T Tauri stars (TTS) is a main step towards the
understanding of star formation in the Galaxy and early phases of
low-mass stellar evolution. Early definitions (e.g. Herbig 1962,
Bastian \etl\ 1983) describe TTS as low-mass pre-main sequence (PMS)
objects showing in their spectrum emission from the hydrogen Balmer
lines and the Ca~{\sc ii} H and K lines, reflecting the fact that TTS
were discovered in \ha\ surveys of nearby molecular clouds
(``classical'' TTS, cTTS).

After the  Einstein Observatory (EO) X-ray mission other objects have
joined the T Tauri family, in particular X-ray active ``weak-line'' TTS
(wTTS), which lack strong emission lines ($W_{\lambda}$(\ha) $\le
10$\AA ) (e.g. Gahm 1980, Feigelson \& DeCampli 1981) and, usually,
near-infrared excess emission (Walter 1986). From X-ray emission
variability and the spatial incompletness of EO pointed observations,
Walter \etl\ (1988) deduced that there should be as many as 1000 wTTS
in the general Taurus-Auriga region, many of which can be discovered
with the RASS. The population defined by the new wTTS discovered by
Walter \etl\ (1988) by optical follow-up observations of previously
unidentified EO sources cannot be identified with  the post-TTS
population proposed by Herbig (1978), as many wTTS share the same locus
in the H-R diagram as cTTS.   In the optical spectrum of all TTS the
\li~absorption doublet is a prominent feature indicative of their PMS
nature (Magazz\`u \etl\ 1992, Mart\'\i n \etl\ 1994).  The majority of
TTS have been discovered in areas where molecular gas has been
detected.  One of the best studied areas is the Taurus-Auriga star
forming region (SFR), a T~association at  a distance of  $\sim 140$~pc
(Elias 1978, Kenyon \etl\ 1994).

The advent of the ROSAT All-Sky Survey (RASS) has enabled us to extend
the search for X-ray active low-mass stars to the complete sky, with a
flux limit comparable with typical EO pointed observations.  By
studying X-ray spectra of RASS-detected well-known TTS and unidentified
RASS sources in the Taurus-Auriga SFR, Neuh\"auser \etl\ (1995a) have
proposed that several hundreds of coronally active TTS are hidden in
the RASS database.  So far, 76 PMS stars have been discovered in the
central parts of the Taurus-Auriga association (Wichmann \etl\ 1996).
Some other PMS stars have been discovered with optical follow-up
observations of sources found in deep ROSAT pointed observations in
Taurus-Auriga (Strom \& Strom 1994, Carkner \etl\ 1996, Wichmann
\etl\ 1996).  Based on ROSAT observations, many new PMS stars have been
discovered in other SFRs as well (see Alcal\'a \etl\ 1995, Neuh\"auser
1996,  Krautter 1996).

As the set of unidentified RASS sources is very large, Sterzik
\etl\ (1995)   proposed an efficient way of pre-selecting TTS
candidates just from ROSAT and Hubble Space Telescope Guide Star
Catalog (GSC) data alone,  which are easily accessible for all
unidentified RASS sources having a near-by GSC counterpart.
Surprisingly, Sterzik \etl\ (1995) did not find any gradient in the
space density of TTS candidates at the edges of the Orion molecular
clouds and they proposed to extend the search for TTS even outside the
cloud complexes.  The strength of RASS as being spatially unbiased
allows us to survey for TTS also outside the regions previously known
to be populated by TTS.

In this paper, we report on the results of such a study conducted south
of the Taurus-Auriga clouds.  In Sect. 2, we describe ROSAT
observations and X-ray data analysis. The method introduced by Sterzik
\etl\ (1995) has been then applied to select TTS candidates (Sect.~3).
Optical follow-up observations of these TTS candidates and data
reduction are explained in Sect.~4, results are presented in Sect.~5
and discussed in the last section.

In addition to data presented here, high-resolution spectral
observations have been obtained for most of our stars to study their
radial velocity; these data as well as proper motions of several new
PMS stars are given in Neuh\"auser \etl\ (1997), in which the
kinematics and possible modes of origin of newly discovered PMS stars
are discussed.  Preliminary results of our study were reported by
Neuh\"auser \etl\ (1995c), who discussed the results of optical
observations of 15 RASS-selected TTS candidates south of Taurus-Auriga.
These objects are included in the sample discussed here.

\section{X-ray data}

The X-ray telescope ROSAT and its detectors are described in detail by
Tr\"umper (1983) and Pfeffermann \etl\ (1988). The Position Sensitive
Proportional Counter (PSPC) on board ROSAT performed an All-Sky Survey
scanning the sky in great circles with a $2^{\circ}$ diameter field of
view. Observations in our selected area have been performed in August
and September 1990 and in February 1991.  Vignetting corrected RASS
exposure times vary roughly with $1/\cos \beta$ where $\beta$ denotes
ecliptic latitude.  In our studied area, the exposure varies between
$\sim 430$ and $\sim 580$ seconds.  During the RASS, all areas in the
sky have been observed in $\sim 30$ second long scans, separated by
$\sim 90$ minutes and spread over almost two days.  The ROSAT PSPC has
$255$ instrumental energy channels sensitive from 0.1~keV to 2.4~keV
(broad band). The spectral resolution of the ROSAT PSPC instrument ($43
\%$ at 0.93~keV) permits a reasonable spectral analysis in three energy
bands:

\begin{itemize}
\item soft: 0.1 to 0.4~keV

\item hard~1: 0.5 to 0.9~keV

\item hard~2: 0.9 to 2.1~keV
\end{itemize}

In the flux-limited RASS, most of the previously known wTTS but only
few cTTS have been detected (Neuh\"auser \etl\ 1995b). It has been
shown (Neuh\"auser \etl\ 1995a, 1995b) that wTTS and cTTS can be
discriminated by their X-ray spectral hardness ratios.  If $Z_{s}$,
$Z_{h1}$, and $Z_{h2}$ denote count rates for the ROSAT energy channels
soft, hard~1, and hard~2, respectively, we define two hardness ratios
as follows:
\begin{displaymath}
HR\,1 = \frac{Z_{h1} + Z_{h2} - Z_{s}}{Z_{h1} + Z_{h2} + Z_{s}}; \quad
HR\,2 = \frac{Z_{h2} - Z_{h1}}{Z_{h2} + Z_{h1}}.
\end{displaymath}

\begin{figure}[t]
\centerline{\psfig {file=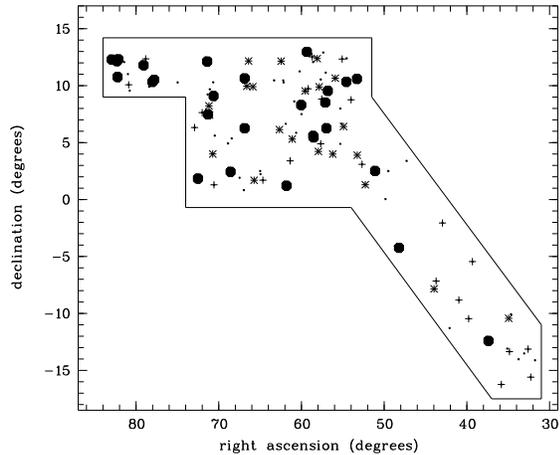,bbllx=0pt,bblly=0pt,bburx=780pt,bbury=539pt,height=7.2cm}}
\caption[]{Taurus-Auriga: In our studied area (enclosed by a box) we find
new PMS stars (filled symbols), possible new PMS stars (stars), new
dKe/dMe stars (plusses) and other non-PMS stars (dots)}
\end{figure}

We have extracted from the merged photon event, exposure, and attitude
files of the RASS data those data which pertain to our studied area
south of the Taurus-Auriga dark clouds. This area is shown in Fig.~1;
it includes a region just south of the Taurus-Auriga dark cloud complex
(i.e. with $\delta \le 14 ^{\circ}$) and an attached strip
perpendicular to the galactic plane.  Neuh\"auser \etl\ (1995c) refer
to the same area, while the region north of $\delta = 14^{\circ}$ has
been searched for new TTS by Wichmann \etl\ (1996).  Source detection
and local background determination were done separately in five
different ROSAT energy bands: broad, soft, hard~1, hard~2, and hard
(0.5 to 2.1~keV). Detected sources were merged and tested again with a
maximum likelihood technique. Details on   RASS data reduction and
source detection algorithm can be found in Neuh\"auser \etl\ (1995b).
These authors showed that all sources with a maximum likelihood of
existence of at least 7.4 can be accepted as real sources; such a low
likelihood value is justified as we search for unknown TTS which may be
X-ray faint.

\section {Source selection}

Out of the $1084$ RASS sources in our studied region,  only $179$
(about $17\%$) can be identified with known stellar and extragalactic
counterparts in the Simbad database. Clearly, for efficient
ground-based optical follow-up observations a pre-selection among the
remaining sample of more than $900$ X-ray sources is needed.  A general
statistical discrimination method was introduced by Sterzik \etl\
(1995) in order to find promising TTS candidates in the ROSAT database
and was applied to study the spatial distribution of TTS candidates in
the Orion SFR. In the present case, we apply the same method to our
sample of RASS sources.  A detailed description of the procedure can be
found  in Sterzik \etl\ (1995). Here we only summarize the relevant
steps:
\begin{enumerate}

\item We establish a ``training set'', consisting of members of
previously classified groups, in our case TTS and non-TTS. As our
observations are carried out close to the Taurus SFR, we conveniently
draw the sources of the training set from follow-up observations of
ROSAT sources reported by Wichmann \etl\ (1996). We use the properties
of $63$ TTS and $87$ non-TTS of their survey (all detected in the
RASS).  We added also $54$ RASS-detected TTS from the Herbig \& Bell
(1988) catalog. Our training set presents a uniform spatial
distribution.

\item As discrimination parameters we choose: $HR\,1$ and $HR\,2$
determined for each X-ray source, $V$ of the closest optical
counterpart in the GSC within $40\arcsec$ around the X-ray source, and
the X-ray to optical flux ratio $f_{X}/f_{V}$.  If no GSC counterpart
is present within $40\arcsec$  we assign $V=20$, indicating a faint
optical X-ray source not detected in the GSC.

\item A $k$-nearest neighbourhood  analysis is applied to all sources
in the sample. A discrimination probability $P_i$ for each source $i$
is defined as \begin{displaymath} P_i = \frac{n(\mbox{TTS})}{k},
\end{displaymath} where $n(\mbox{TTS})$ is the number of TTS in the
training set ``in the neighbourhood'' of source $i$ in the
four-dimensional discrimination parameter space. The neighbourhood is
defined as the sphere containing exactly
\begin{displaymath}
n(\mbox{TTS}) + n(\mbox{non-TTS}) = k
\end{displaymath}
members of the training set. For the present study we have chosen
$k=16$, because this value proved to give the best results in telling
TTS from non-TTS within the training set. High values of $P_i$ indicate
that the source properties are similar to those in the sample of TTS of
the training set.
\end{enumerate}

Source selection on this basis implies a bias towards (X-ray and
optical) properties of typical TTS in the training set, as desired.
However, the selection cannot guarantee a ``pure'' TTS sample, because
other classes of stars such as active binaries, RS~CVn  binaries,
emission line dwarfs, etc., share similar values within the
discrimination parameter set, and are expected to contaminate any
sub-sample.

Figure 2 summarizes the definition of the X-ray sample according to the
discrimination probability $P$. We note that, above a discrimination
threshold of 0.6 (0.5), we have observed 97 (110) sources out of a
total of 180 (246) candidates.  It can be seen that many X-ray sources
with high discrimination probability, i.e. many TTS candidates, indeed
are new PMS stars (see Sects.~4,5), while there are no new PMS stars
among X-ray sources with low discrimination probability.

\begin{figure}
\centerline{\psfig {file=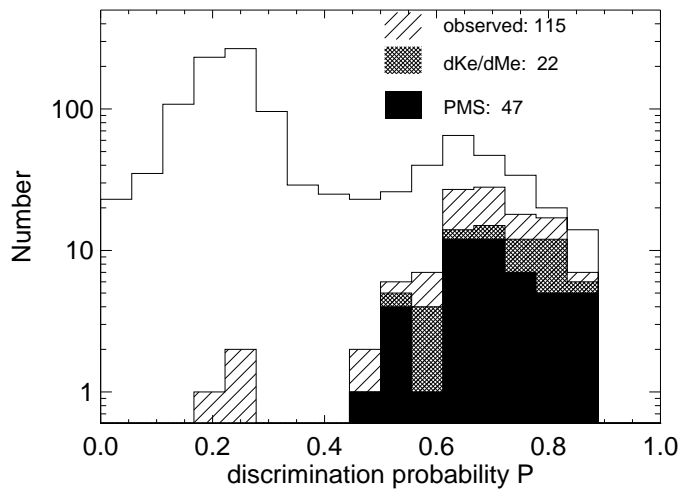,bbllx=38pt,bblly=18pt,bburx=663pt,bbury=474pt,height=6.5cm}}
\caption[]{Source selection:  The histogram gives the distribution of
discrimination probability $P$ for all 1084 X-ray sources in our
studied area.  Plotted is the number of stars per $P$ bin (with bin
size 0.05).  The hatched histogram indicates the observed sub-sample
with dKe/dMe stars and new PMS stars (including certain and possible
new PMS stars).  There are a total of 47 X-ray counterparts classified
as either  PMS or PMS? and 22 other X-ray counterparts classified as
dKe or dMe stars. There are no RASS sources with $P > 0.9$}
\end{figure}

Whereas previously this classification scheme was primarily applied to
demonstrate the discrimination ability within the training set itself
(as the only  a priori  known classified sample), the results of this
work indeed prove the prediction quality of the suggested procedure and
allow a reliable extrapolation of the number of different types of
active stars in the sample.

In Table~1 we list all the RASS sources selected in the way described
above.  We list ROSAT source name, X-ray counts (background subtracted
and vignetting corrected) detected in the broad ROSAT band, exposure
time, hardness ratios, the maximum likelihood for existence of the
source, and spectral fit results.  Spectral fits are performed -- as
described in Neuh\"auser \etl\ (1995b) -- using the hardness ratios and
assuming a one-temperature Raymond-Smith model (Raymond \& Smith
1977).  Both the emission energy k$T_{\rm X}$ and the absorbing
foreground column density $N_{\rm H}$ are taken as free fit parameters.
The procedure searches for the (k$T_{\rm X}$, $N_{\rm H}$) pair that
best fits the observed ($HR\,1$, $HR\,2$) pair. From the ROSAT PSPC
response matrix and the fit results, we can then compute the
individually spectral-corrected X-ray flux (also given in Table 1).
The X-ray source positions can be obtained from the positions of the
optical counterparts and the offsets between X-ray and optical
positions (given in Table 2).
\begin{table*}
\caption []{X-ray data for our sample. Listed are all X-ray sources
detected in the ROSAT All-Sky Survey (sorted by right ascension),
studied optically in this work. We list ROSAT source designation (X-ray
positions can be obtained from data in Table 2), number of  counts in
the broad ROSAT band with errors, exposure time, hardness ratios with
errors, the maximum likelihood of existence $ML$, the emission energy
(typical error being $\pm 0.55$~keV), the absorbing foreground column
density (typical error $\pm 0.77$, in $\log (N_{\rm H})~{\rm
cm}^{-2})$, and finally the X-ray flux (typical error $\pm 1.87
\,10^{-13} \mbox {erg s}^{-1} {\rm cm}^{-2}$)}

\begin {flushleft}
\begin{tabular}{lr@{$\pm$}lcr@{$\pm$}lr@{$\pm$}lrccrc} 
\noalign{\smallskip}
\hline
\noalign{\smallskip}
  designation & \multicolumn{2}{c}{counts} & exp  & \multicolumn{2}{c}{$HR\,1$} & \multicolumn{2}{c}{$HR\,2$} 
  & $ML$  & k$ T_{\rm X}$   & $\log (N_{\rm H})$  & {$f_{\rm X}/10^{-13}$} \hspace {-0.5 cm}   &\\
~ & \multicolumn{2}{c}{~} &  (s)  & \multicolumn{2}{c}{~} & \multicolumn{2}{c}{~}  &  &  (keV)  &  (cm$^{-2})$ &  (erg~s$^{-1}$~cm$^{-2}$) \hspace {-0.7 cm} & \hspace {0.5 cm}\\ 
\noalign{\smallskip}
\hline
\noalign{\smallskip}
RXJ0207.0$-$1407 &    7.1 & 3.3 &442.1 & 1.00 &0.21 & 0.69 &0.31 & 14.3 &  1.14 & 22.00 & 19.56 &\\
RXJ0209.1$-$1536 &   16.3 & 5.5 &444.2 & 0.95 &0.49 &-0.09 &0.35 & 21.8 &  0.72 & 20.85 &  5.17 &\\
RXJ0210.4$-$1308 &   21.5 & 7.3 &548.2 & 0.35 &0.31 & 0.20 &0.39 & 17.7 &  1.07 & 18.10 &  3.53 &\\
RXJ0212.3$-$1330 &   17.9 & 6.8 &556.2 & 0.20 &0.38 & 0.23 &0.45 & 10.9 &  1.22 & 18.00 &  2.98 &\\
RXJ0215.0$-$1402 &    9.9 & 4.5 &529.3 & 0.67 &0.38 & 0.43 &0.53 & 12.4 &  1.67 & 20.35 &  3.06 &\\
RXJ0218.6$-$1004 &   17.9 & 5.6 &486.2 & 1.00 &0.12 &-0.04 &0.31 & 21.7 &  0.58 & 21.55 & 11.53 &\\
RXJ0219.4$-$1321 &   11.6 & 4.7 &523.8 & 0.92 &0.63 &-0.03 &0.40 & 12.3 &  0.81 & 20.60 &  2.76 &\\
RXJ0219.7$-$1026 &   46.2 & 9.4 &530.5 & 0.49 &0.19 & 0.46 &0.19 & 53.1 &  1.70 & 20.10 & 12.16 &\\
RXJ0220.4$-$1305 &   15.8 & 6.5 &506.5 & 0.07 &0.45 &-0.23 &0.25 & 10.5 &  0.39 & 18.00 &  2.31 &\\
RXJ0223.3$-$1615 &    8.6 & 4.1 &532.1 & 1.00 &0.22 & 0.18 &0.46 &  9.6 &  0.57 & 21.85 & 21.09 &\\
RXJ0229.5$-$1224 &   12.3 & 5.0 &516.4 & 0.60 &0.35 & 0.41 &0.37 & 12.8 &  1.68 & 20.25 &  3.70 &\\
RXJ0237.3$-$0527 &    9.2 & 4.5 &279.4 & 0.22 &0.48 & 0.44 &0.68 &  7.7 &  1.63 & 19.45 &  3.60 &\\
RXJ0239.1$-$1028 &    6.5 & 3.3 &326.0 & 0.33 &0.51 & 0.26 &0.24 & 12.7 &  1.15 & 18.90 &  1.91 &\\
RXJ0243.9$-$0850 &    8.3 & 4.5 &341.7 & 0.44 &0.44 & 0.55 &0.74 &  7.5 &  1.67 & 20.00 &  3.22 &\\
RXJ0248.3$-$1117 &    8.0 & 4.3 &418.7 & 0.16 &0.28 & 0.01 &0.38 &  7.7 &  1.15 & 18.00 &  1.76 &\\
RXJ0251.8$-$0203 &    8.5 & 3.9 &325.7 & 0.43 &0.43 & 0.09 &0.46 & 12.6 &  0.96 & 18.00 &  2.28 &\\
RXJ0254.8$-$0709 &   10.4 & 4.4 &329.7 & 0.70 &0.32 & 0.76 &0.26 & 15.7 &  1.94 & 21.80 & 20.03 &\\
RXJ0255.8$-$0750 &    8.7 & 4.3 &330.5 & 0.89 &0.80 & 0.30 &0.42 & 10.0 &  1.17 & 20.60 &  4.15 &\\
RXJ0309.1+0324 &   28.5 & 7.2 &614.3 & 0.81 &0.19 & 0.39 &0.24 & 38.1 &  1.39 & 20.50 &  7.75 &\\
RXJ0312.8$-$0414 &   34.0 & 7.9 &424.7 & 0.64 &0.20 & 0.07 &0.23 & 43.4 &  0.93 & 19.70 &  8.31 &\\
RXJ0314.8$-$0406 &    6.7 & 3.9 &465.8 & 1.00 &0.29 & 0.46 &0.48 &  8.5 &  0.76 & 22.00 & 24.47 &\\
RXJ0317.9+0231 &   14.3 & 5.6 &541.9 & 1.00 &0.16 &-0.01 &0.39 & 12.2 &  0.58 & 21.60 & 10.67 &\\
RXJ0319.3+0003 &    8.7 & 4.4 &603.2 & 1.00 &0.26 &-0.10 &0.51 &  8.8 &  0.55 & 21.45 &  4.66 &\\
RXJ0324.4+0231 &   21.7 & 7.1 &461.1 & 0.36 &0.34 & 0.15 &0.38 & 16.6 &  1.03 & 18.00 &  4.23 &\\
RXJ0329.1+0118 &   11.9 & 4.9 &637.8 & 1.00 &0.14 & 0.50 &0.35 & 12.5 &  0.83 & 22.00 & 28.73 &\\
RXJ0330.7+0306 &   54.4 & 9.8 &737.7 & 1.00 &0.04 & 0.05 &0.18 & 54.9 &  0.59 & 21.70 & 40.64 &\\
RXJ0333.0+0354 &    9.9 & 4.5 &634.2 & 1.00 &0.22 & 0.57 &0.46 & 10.5 &  0.94 & 22.00 & 21.98 &\\
RXJ0333.1+1036 &   20.3 & 6.2 &564.3 & 0.36 &0.25 & 0.48 &0.26 & 30.8 &  1.66 & 19.85 &  4.55 &\\
RXJ0336.0+0846 &    8.7 & 4.0 &543.0 & 0.78 &0.85 &-0.26 &0.44 & 10.7 &  0.51 & 20.30 &  1.72 &\\
RXJ0338.1+1224 &    8.3 & 4.2 &555.1 & 1.00 &0.23 & 0.25 &0.50 & 10.0 &  0.60 & 21.90 & 18.40 &\\
RXJ0338.3+1020 &   13.2 & 4.8 &529.8 & 1.00 &0.16 & 0.17 &0.35 & 15.1 &  0.64 & 21.80 & 19.04 &\\
RXJ0339.6+0624 &    8.6 & 4.1 &569.3 & 1.00 &0.20 & 0.47 &0.40 &  9.7 &  0.78 & 22.00 & 24.42 &\\
RXJ0340.3+1220 &   26.9 & 6.6 &517.6 & 0.85 &0.17 &-0.18 &0.25 & 49.3 &  0.64 & 20.35 &  5.77 &\\
RXJ0340.5+0639 &    5.3 & 2.9 &522.8 & 1.00 &0.26 & 0.26 &0.25 &  9.8 &  0.61 & 21.90 & 12.48 &\\
RXJ0341.2+0453 &   37.6 & 8.2 &538.1 & 1.00 &0.04 & 0.26 &0.21 & 56.6 &  0.63 & 21.90 & 81.67 &\\
RXJ0341.2+0759 &    8.1 & 3.9 &567.9 & 1.00 &0.23 & 0.19 &0.47 &  8.3 &  0.59 & 21.85 & 17.56 &\\
RXJ0343.6+1039 &   17.5 & 6.0 &555.3 & 0.89 &0.30 & 0.60 &0.24 & 21.0 &  2.02 & 21.45 & 11.78 &\\
RXJ0344.8+0359 &   25.5 & 6.9 &529.7 & 0.88 &0.13 &-0.15 &0.29 & 29.0 &  0.67 & 20.45 &  5.55 &\\
RXJ0347.2+0933 &   16.1 & 5.4 &527.6 & 0.58 &0.28 & 0.85 &0.16 & 19.9 &  2.02 & 22.00 & 32.54 &\\
RXJ0347.9+0616 &    8.6 & 4.0 &466.0 & 1.00 &0.24 & 0.43 &0.19 & 11.3 &  0.70 & 22.00 & 33.14 &\\
RXJ0348.2+1109 &   23.8 & 6.5 &471.2 & 0.30 &0.28 &-0.07 &0.32 & 29.7 &  0.93 & 18.00 &  4.41 &\\
RXJ0348.5+0832 &   23.3 & 6.4 &486.9 & 0.92 &0.23 &-0.28 &0.29 & 30.6 &  0.44 & 20.80 &  6.81 &\\
RXJ0349.4+1255 &   19.7 & 5.7 &479.3 & 0.81 &0.35 &-0.25 &0.29 & 42.5 &  0.53 & 20.35 &  4.56 &\\
RXJ0350.2+0849 &    6.6 & 3.3 &470.2 & 1.00 &0.24 & 0.07 &0.51 & 11.8 &  0.61 & 21.70 &  7.74 &\\
RXJ0350.6+0454 &    7.4 & 3.7 &488.1 & 1.00 &0.23 & 0.09 &0.50 & 11.6 &  0.58 & 21.75 & 12.09 &\\
RXJ0350.6+1033 &   12.5 & 5.3 &569.5 & 1.00 &0.17 & 0.51 &0.33 &  9.4 &  0.85 & 22.00 & 33.80 &\\
RXJ0351.4+0953 &   11.8 & 4.9 &491.4 & 0.90 &0.24 &-0.25 &0.42 &  8.8 &  0.51 & 20.65 &  2.95 &\\
RXJ0351.8+0413 &    6.5 & 3.3 &441.1 & 1.00 &0.25 &-0.10 &0.52 &  9.2 &  0.55 & 21.45 &  4.77 &\\
RXJ0352.4+1223 &   11.8 & 4.8 &522.4 & 1.00 &0.18 &-0.00 &0.42 & 13.3 &  0.55 & 21.65 &  9.48 &\\
RXJ0354.1+0528 &    7.4 & 4.1 &535.0 & 1.00 &0.26 & 0.55 &0.19 &  7.6 &  0.91 & 22.00 & 20.34 &\\
\noalign{\smallskip}
 \hline
\end{tabular}
\end {flushleft}
\end{table*}

\addtocounter{table}{-1}

\begin{table*}

\caption []{Continued}
\begin{flushleft}
\begin{tabular}{lr@{$\pm$}lcr@{$\pm$}lr@{$\pm$}lrccrc} 
\noalign{\smallskip}
\hline
\noalign{\smallskip}
  designation & \multicolumn{2}{c}{counts} & exp  & \multicolumn{2}{c}{$HR\,1$} & \multicolumn{2}{c}{$HR\,2$} 
  & $ML$  & k$ T_{\rm X}$   & $\log (N_{\rm H})$  & {$f_{\rm X}/10^{-13}$} \hspace {-0.5 cm}   &\\
~ & \multicolumn{2}{c}{~} &  (s)  & \multicolumn{2}{c}{~} & \multicolumn{2}{c}{~}  &  &  (keV)  &  (cm$^{-2})$ &  (erg~s$^{-1}$~cm$^{-2}$) \hspace {-0.7 cm} & \hspace {0.5 cm}\\ 
\noalign{\smallskip}
\hline
\noalign{\smallskip}
RXJ0354.3+0535 &   19.2 & 6.1 &536.7 & 1.00 &0.11 & 0.42 &0.31 & 24.7 &  0.69 & 22.00 & 64.23 &\\
RXJ0354.4+1204 &    8.2 & 3.9 &474.2 & 0.58 &0.39 & 1.00 &1.00 & 13.2 &  1.62 & 20.20 &  2.53 &\\
RXJ0354.8+1232 &   16.6 & 5.9 &505.8 & 0.81 &0.33 &-0.11 &0.36 & 18.3 &  0.73 & 20.15 &  3.42 &\\
RXJ0356.7+0943 &   10.3 & 5.1 &507.5 & 0.60 &0.39 &-0.40 &0.42 &  9.1 &  0.32 & 20.35 &  2.80 &\\
RXJ0357.3+1258 &   17.9 & 6.1 &486.0 & 0.65 &0.24 &-0.39 &0.30 & 15.9 &  0.33 & 20.40 &  4.63 &\\
RXJ0358.1+0932 &   27.6 & 6.9 &467.6 & 0.86 &0.22 & 0.57 &0.20 & 42.7 &  2.02 & 21.35 & 19.43 &\\
RXJ0400.0+0730 &   10.0 & 4.1 &459.9 & 1.00 &0.15 & 0.61 &0.33 & 14.9 &  1.00 & 22.00 & 29.38 &\\
RXJ0400.1+0818 &  169.6 &16.6 &450.7 & 0.27 &0.10 & 0.17 &0.12 &298.5 &  1.11 & 18.00 & 34.22 &\\
RXJ0400.8+1116 &   15.7 & 5.7 &494.8 & 1.00 &0.15 &-0.05 &0.38 & 13.4 &  0.56 & 21.55 & 10.26 &\\
RXJ0402.5+0552 &   18.2 & 5.8 &414.1 & 0.68 &0.27 & 0.24 &0.37 & 19.5 &  1.11 & 20.05 &  5.37 &\\
RXJ0403.5+0837 &   20.2 & 6.2 &480.6 & 1.00 &0.09 & 0.56 &0.25 & 22.3 &  0.93 & 22.00 & 59.17 &\\
RXJ0404.4+0519 &   20.3 & 5.6 &413.6 & 1.00 &0.06 &-0.06 &0.28 & 41.8 &  0.57 & 21.55 & 15.87 &\\
RXJ0405.5+0324 &   34.2 & 7.8 &445.4 & 1.00 &0.04 & 0.03 &0.23 & 54.4 &  0.57 & 21.70 & 44.18 &\\
RXJ0407.2+0113 &   10.1 & 4.1 &421.0 & 1.00 &0.16 & 0.35 &0.39 & 13.0 &  0.52 & 22.00 & 56.03 &\\
RXJ0407.6+0638 &   19.2 & 6.4 &474.5 & 1.00 &0.08 &-0.25 &0.34 & 16.3 &  0.46 & 21.30 &  9.69 &\\
RXJ0408.6+1017 &    5.9 & 3.4 &375.6 & 1.00 &0.25 & 0.53 &0.43 &  8.0 &  0.88 & 22.00 & 23.10 &\\
RXJ0408.8+1028 &  106.0 &13.6 &387.9 & 0.52 &0.11 & 0.24 &0.14 &159.7 &  1.11 & 19.65 & 29.51 &\\
RXJ0409.8+1209 &   10.5 & 5.1 &357.3 & 0.76 &0.65 & 0.30 &0.23 &  7.7 &  1.19 & 20.30 &  4.21 &\\
RXJ0410.6+0608 &    8.7 & 3.9 &395.1 & 1.00 &0.19 & 0.08 &0.46 & 10.1 &  0.56 & 21.75 & 18.47 &\\
RXJ0413.2+1028 &   13.0 & 4.6 &374.5 & 1.00 &0.01 & 0.56 &0.11 & 20.0 &  0.97 & 22.00 & 48.87 &\\
RXJ0418.6+0143 &    9.2 & 4.5 &411.3 & 0.52 &0.40 & 0.46 &0.39 &  8.2 &  1.61 & 20.10 &  3.11 &\\
RXJ0419.8+0214 &   17.3 & 6.1 &404.5 & 1.00 &0.12 & 0.11 &0.35 & 13.6 &  0.62 & 21.75 & 34.11 &\\   
RXJ0419.9+0231 &   47.7 & 9.7 &385.4 & 1.00 &0.04 & 0.06 &0.18 & 12.5 &  0.61 & 21.70 &  7.69 &\\ 
RXJ0422.9+0141 &   15.4 & 5.9 &444.8 & 0.90 &0.25 & 0.50 &0.33 & 10.1 &  1.78 & 20.85 &  7.51 &\\
RXJ0423.5+0955 &   33.9 & 8.8 &577.2 & 0.01 &0.26 &-0.00 &0.35 & 22.9 &  2.02 & 18.00 &  5.54 &\\
RXJ0425.5+1210 &   21.3 & 6.2 &641.0 & 1.00 &0.08 &-0.09 &0.29 & 23.5 &  0.55 & 21.50 & 10.75 &\\
RXJ0426.4+0957 &   16.3 & 5.6 &657.7 & 1.00 &0.12 &-0.09 &0.34 & 18.6 &  0.54 & 21.50 &  8.01 &\\
RXJ0427.4+1039 &   20.6 & 6.3 &654.3 & 1.00 &0.10 & 0.62 &0.25 & 23.3 &  1.02 & 22.00 & 42.55 &\\
RXJ0427.5+0616 &   67.2 &11.4 &612.2 & 0.37 &0.16 &-0.06 &0.20 & 75.5 &  0.90 & 18.00 &  9.45 &\\
RXJ0427.8+0049 &   17.8 & 6.0 &495.9 & 0.84 &0.23 &-0.05 &0.34 & 17.7 &  0.80 & 20.25 &  4.04 &\\
RXJ0429.9+0155 &   15.1 & 5.1 &534.7 & 1.00 &0.11 & 0.33 &0.30 & 18.5 &  0.63 & 21.95 & 53.69 &\\
RXJ0433.7+0522 &   15.1 & 6.3 &817.3 & 1.00 &0.18 & 0.31 &0.39 & 10.2 &  0.59 & 21.95 & 37.34 &\\
RXJ0434.3+0226 &   33.6 & 8.1 &775.6 & 1.00 &0.04 & 0.11 &0.24 & 39.0 &  0.56 & 21.80 & 36.33 &\\
RXJ0435.5+0455 &   19.6 & 6.0 &746.2 & 1.00 &0.08 & 0.37 &0.26 & 25.4 &  0.58 & 22.00 & 53.09 &\\
RXJ0441.9+0537 &   96.3 &13.6 &582.9 & 1.00 &0.01 & 0.50 &0.12 &123.1 &  0.89 & 22.00 &242.90 &\\
RXJ0442.3+0118 &   19.8 & 7.1 &528.9 & 0.27 &0.44 & 0.34 &0.43 & 14.1 &  1.31 & 19.15 &  3.76 &\\
RXJ0442.5+0906 &   22.0 & 6.4 &532.0 & 0.36 &0.28 &-0.03 &0.29 & 24.2 &  0.91 & 18.00 &  3.56 &\\
RXJ0442.6+1018 &   83.9 &12.0 &537.8 & 0.90 &0.07 & 0.07 &0.14 &142.0 &  0.91 & 20.55 & 19.83 &\\
RXJ0442.9+0400 &   17.4 & 5.9 &532.6 & 1.00 &0.11 & 0.11 &0.34 & 21.3 &  0.62 & 21.75 & 26.06 &\\
RXJ0444.3+0941 &   62.2 &10.4 &568.1 & 0.58 &0.14 & 0.11 &0.17 & 95.5 &  0.97 & 19.55 & 10.84 &\\
RXJ0444.4+0725 &   12.6 & 4.9 &514.6 & 1.00 &0.15 &-0.30 &0.39 & 12.6 &  0.35 & 21.30 &  7.16 &\\
RXJ0444.7+0814 &   18.9 & 6.6 &571.0 & 0.58 &0.33 & 0.32 &0.35 & 19.2 &  1.25 & 20.00 &  4.27 &\\
RXJ0445.2+0729 &   10.1 & 5.0 &568.8 & 1.00 &0.21 & 0.19 &0.46 &  8.0 &  0.59 & 21.85 & 21.86 &\\
RXJ0445.3+0914 &    9.5 & 4.7 &600.6 & 1.00 &0.21 & 0.43 &0.51 &  8.9 &  0.70 & 22.00 & 28.40 &\\
RXJ0445.5+1207 &    9.4 & 4.4 &617.3 & 0.78 &0.50 & 0.60 &0.40 & 11.6 &  2.02 & 21.40 &  5.01 &\\
RXJ0448.0+0738 &   88.8 &12.8 &535.0 & 0.28 &0.14 &-0.13 &0.18 & 93.9 &  0.44 & 18.00 & 12.52 &\\
RXJ0450.0+0151 &   20.6 & 6.3 &535.6 & 0.95 &0.21 &-0.08 &0.31 & 21.1 &  0.74 & 20.80 &  5.16 &\\
RXJ0451.6+0619 &    9.5 & 4.3 &558.8 & 1.00 &0.18 &-0.04 &0.45 &  8.9 &  0.58 & 21.55 &  5.33 &\\
RXJ0459.9+1017 &   29.4 & 7.7 &603.5 & 1.00 &0.05 & 0.32 &0.23 & 27.8 &  0.62 & 21.95 & 98.47 &\\
RXJ0511.2+1031 &   18.4 & 6.4 &661.5 & 1.00 &0.13 & 0.26 &0.37 & 15.7 &  0.62 & 21.90 & 34.24 &\\
RXJ0511.9+1112 &   12.9 & 4.8 &618.1 & 0.71 &0.36 & 0.29 &0.37 & 12.8 &  1.18 & 20.20 &  2.87 &\\
RXJ0512.0+1020 &   23.6 & 6.7 &651.2 & 0.78 &0.21 & 0.13 &0.29 & 23.4 &  0.99 & 20.20 &  4.46 &\\
RXJ0513.6+0955 &    9.2 & 3.9 &593.2 & 0.43 &0.42 &-0.45 &0.17 & 14.4 &  0.29 & 20.25 &  1.98 &\\
\noalign{\smallskip}
\hline

\end{tabular}
\end {flushleft}
\end{table*}

\addtocounter{table}{-1}
\begin{table*}

\caption []{Continued}

\begin {flushleft}
\begin{tabular}{lr@{$\pm$}lcr@{$\pm$}lr@{$\pm$}lrccrc} 
\noalign{\smallskip}
\hline
\noalign{\smallskip}
  designation & \multicolumn{2}{c}{counts} & exp  & \multicolumn{2}{c}{$HR\,1$} & \multicolumn{2}{c}{$HR\,2$} 
  & $ML$  & k$ T_{\rm X}$   & $\log (N_{\rm H})$  & {$f_{\rm X}/10^{-13}$} \hspace {-0.5 cm}   &\\
~ & \multicolumn{2}{c}{~} &  (s)  & \multicolumn{2}{c}{~} & \multicolumn{2}{c}{~}  &  &  (keV)  &  (cm$^{-2})$ &  (erg~s$^{-1}$~cm$^{-2}$) \hspace {-0.7 cm} & \hspace {0.5 cm}\\
\noalign{\smallskip}
 \hline
\noalign{\smallskip}
RXJ0515.3+1221 &   27.5 & 7.9 &690.5 & 0.57 &0.24 &-0.07 &0.30 & 24.4 &  0.79 & 19.05 &  3.46 &\\ 
RXJ0516.3+1148 &   24.9 & 7.3 &658.6 & 0.98 &0.24 &-0.03 &0.29 & 23.8 &  0.54 & 21.60 & 15.87 &\\
RXJ0523.0+0934 &    8.3 & 4.1 &429.6 & 0.74 &0.52 &-0.25 &0.51 &  8.0 &  0.54 & 20.15 &  1.92 &\\
RXJ0523.5+1005 &   28.7 & 6.9 &466.5 & 0.93 &0.15 & 0.15 &0.22 & 49.7 &  0.98 & 20.70 &  9.10 &\\
RXJ0523.9+1101 &    6.8 & 3.3 &498.8 & 0.28 &0.40 &-0.10 &0.24 & 16.2 &  0.78 & 18.00 &  1.14 &\\
RXJ0525.7+1205 &   15.6 & 6.3 &846.4 & 1.00 &0.15 & 0.62 &0.13 &  9.9 &  1.02 & 22.00 & 24.91 &\\
RXJ0528.4+1213 &   33.8 & 7.9 &661.9 & 1.00 &0.05 & 0.06 &0.23 & 43.3 &  0.61 & 21.70 & 28.14 &\\
RXJ0528.5+1219 &   11.0 & 4.4 &626.0 & 0.66 &0.29 & 0.30 &0.37 & 17.0 &  1.19 & 20.10 &  2.32 &\\
RXJ0528.9+1046 &   13.2 & 5.0 &639.2 & 1.00 &0.16 & 0.64 &0.25 & 21.4 &  1.05 & 22.00 & 26.86 &\\
RXJ0529.3+1210 &   23.2 & 6.9 &631.9 & 0.15 &0.29 & 0.03 &0.42 & 24.2 &  1.20 & 18.00 &  3.40 &\\
RXJ0530.9+1227 &   28.2 & 7.5 &693.4 & 1.00 &0.06 & 0.34 &0.25 & 23.2 &  0.51 & 22.00 & 94.99 &\\
RXJ0531.8+1218 &   10.5 & 4.8 &688.5 & 0.61 &0.30 &-0.02 &0.46 &  7.9 &  0.84 & 19.40 &  1.42 &\\ 
\noalign{\smallskip}
\hline

\end {tabular}
\end {flushleft}
\end {table*}

In Table~2 we list the closest optical counterpart to each X-ray source
which is sufficiently bright to be a potential PMS star at the distance
of Taurus-Auriga (e.g., having $V$ brighter than $\sim 16$ mag).  All
but two of the counterparts listed in Table 2 are GSC stars.  This is a
consequence of the selection process, as we have searched for GSC
counterparts and used their $V$ magnitudes to select PMS candidates.
The two non-GSC stars selected are listed in Simbad but, before our
observations, it was not known whether they were TTS or not.  In some
cases, we have performed optical spectroscopy for several possible
counterparts, when more than one star was found close to the X-ray
position.  As their $V$ magnitudes have not been used to estimate the
likelihood that the relevant X-ray source may be a PMS star, we do not
list these additional counterparts in Table~2, but in Table~3.  If
several optical counterparts are observed, we distinguish them by the
  letters at the end of the relevant ROSAT source designations
(identifying the relative positions of the different counterparts). In
Table 2 we list also the offset between the X-ray and optical positions
(X $-$ optical), the $V$ magnitude, the X-ray to optical flux ratio,
and the discrimination probability $P$, i.e. the value that reflects to
which degree a source resembles typical TTS properties.
 
\section{Optical follow-up observations}

Medium-resolution spectra for the objects of our sample were obtained
with the Intermediate Dispersion Spectrograph at the Cassegrain focus
of the Isaac Newton Telescope in La~Palma. The 500~mm camera, equipped
with a TEK TK1024A CCD, was used in conjunction with the gratings
R600\,R and R1200\,Y, giving a resolution $\lambda/\Delta\lambda$ of
about 4200 and 8400, respectively. The observations were performed in
October 1994, when both gratings were used, and in November 1995, when
only the R1200\,Y was used.  Exposure times ranged from 200 to 1200
seconds. Close to each stellar frame a wavelength calibration CuNe lamp
was exposed; tungsten flat field exposure were taken both at the
beginning and at the end of the night.

The CCD frames were reduced with the IRAF package\footnote{IRAF is
distributed by the National Optical Observatory, which is operated by
the Association of Universities for Research in Astronomy, Inc., under
contract with the National Science Foundation.}.  Each image was
de-biased and flat fielded. The spectra were extracted with the IRAF
task {\tt apall}, which allows optimum extraction, cleaning from cosmic
ray events, and background subtraction. Wavelength calibration was
performed using the dispersion solution determined on the CuNe spectra
by fitting an order two polynomial, with a rms of 0.03~\AA\ for the
R600\,R grating and 0.01~\AA\ for the R1200\,Y grating.

Low-resolution spectra of additional objects were obtained at the
European Southern Observatory (ESO) using the 1.52\,m telescope
equipped with a Boller \& Chivens spectrograph, in November 1995. A 900
grooves/mm (ESO \# 5) grating and the CCD FORD $2048L$ of
$2048~\times~2048$ pixels were used.  With this set-up a mean
resolution of about 2.5 \AA\ (FWHM) in the 4600-7000~\AA\ spectral
range was achieved. The reduction of these spectra was carried out
using the MIDAS package. Bias and dark subtraction was first performed
on each frame. The 2-D  frames were then divided by a mean flat-field
and then calibrated in wavelength. The sky subtraction, the extraction
of one dimensional spectra and the flux calibration using a mean
response function, were finally performed.

Spectral types have been assigned using the library of stellar spectra
by Jacoby \etl\ (1984), available in digital form. For each of our
stars a first guess of the spectral type was given  by visual
comparison. Then, we performed an iterative comparison in the following
way:  each spectrum was first normalized to the continuum; the spectral
type standard spectrum was then normalized and rebinned to the same
resolution and spectral range as the problem star; an overplot of the
two spectra allowed us to compare directly different spectral features
and reject or accept the guessed spectral type. Finally, we assigned
the spectral type  of the  standard which follows most closely the
features of the problem star.  Particular attention was given to
features like Ca\,{\sc i} 6718~\AA, Fe~{\sc i} blend 6495~\AA\ and TiO
bands. The \ha\ line, in emission in most of our stars, was not used
for the spectral classification. Except for the spectra obtained at ESO
(which cover the $4700$ to $6800$~\AA\ spectral range), all of our
spectra cover a short spectral range (from $6300$ to $6800$~\AA).
Bearing this in mind, we estimate that our classification can be
uncertain to about one or two sub-classes and $\pm$3 sub-classes for
the stars earlier than about G5.

\begin{table*}

\caption []{Closest optical counterparts. Listed is the closest optical
counterpart to each X-ray source in Table 1.
 We list designation (as in Table 1, but with indications of the
particular counterpart meant here, if several counterparts have been
observed), optical position (for J\,2000.0) from the GSC (or Simbad if
available), offset between X-ray and optical position, V magnitude from
the GSC (or Simbad as indicated by colons if available), X-ray to
optical flux ratio, discrimination probability $P$, and remarks.  Note
that some GSC counterparts appear on several GSC plates which may have
different colors and different filters; we have always chosen the
closest counterpart, i.e.\ we neither try to average the magnitudes nor
indicate the plate color as GSC magnitudes are very uncertain anyway
(about half a magnitude)}

\begin {flushleft}
\begin{tabular}{lrrrrrccl} 
\noalign {\smallskip}
\hline
\noalign {\smallskip}
  designation & $\alpha$ (opt.) & $\delta$ (opt.) & $\Delta \alpha (\arcsec)$ & $\Delta \delta (\arcsec)$ &
    $V$ & $\log (f_{\rm X}/f_{\rm V})$ & $P$ & remarks \\ 
\noalign {\smallskip}
\hline
\noalign {\smallskip}
 RXJ0207.0-1407  & 2 07 02.94 & -14 06 52.0 & -33.0 &  -5.8 & 15.1 & -0.83 & 0.65 &\\
  RXJ0209.1-1536  & 2 09 06.70 & -15 35 43.0 &  -3.6 &   0.7 & 13.0 & -1.32 & 0.81 &\\
  RXJ0210.4-1308NE& 2 10 25.84 & -13 07 57.0 &   3.0 &   4.9 & 10.5 & -2.29 & 0.76 &\\
  RXJ0212.3-1330  & 2 12 18.72 & -13 30 42.1 &  17.6 & -17.3 & 11.4 & -2.01 & 0.65 &\\
  RXJ0215.0-1402  & 2 15 00.45 & -14 01 31.9 &  13.8 &  31.9 & 15.1 & -0.77 & 0.54 &\\
  RXJ0218.6-1004  & 2 18 39.54 & -10 04 05.0 & -12.3 &   6.3 & 11.8 & -1.79 & 0.81 &\\
  RXJ0219.4-1321B & 2 19 25.30 & -13 21 06.6 &   9.6 & -12.0 & 14.8 & -0.81 & 0.65 &\\
  RXJ0219.7-1026  & 2 19 47.38 & -10 25 39.9 &  -9.3 &   6.9 & 11.6 & -1.50 & 0.70 &\\
  RXJ0220.4-1305  & 2 20 28.34 & -13 05 23.9 &   0.5 & -18.8 & 14.6 & -0.75 & 0.27 &\\
  RXJ0223.3-1615NE& 2 23 21.41 & -16 14 20.9 &  -5.3 &  12.8 & 13.4 & -1.51 & 0.86 &\\
  RXJ0229.5-1224  & 2 29 35.05 & -12 24 08.8 &  13.0 & -13.6 &  9.9:& -2.74 & 0.54 &HD 15526  \\
  RXJ0237.3-0527  & 2 37 19.85 & -05 26 54.1 &   4.6 &   7.6 & 13.3 & -1.24 & 0.54 &\\
  RXJ0239.1-1028  & 2 39 08.72 & -10 27 44.1 &  -0.6 &   2.7 & 13.2 & -1.50 & 0.65 &\\
  RXJ0243.9-0850  & 2 43 52.29 & -08 49 30.3 &  18.3 &  14.2 & 12.9 & -1.53 & 0.59 &\\
  RXJ0248.3-1117  & 2 48 22.16 & -11 17 12.3 &  25.8 &   6.4 & 12.0 & -2.00 & 0.65 &\\
  RXJ0251.8-0203  & 2 51 48.47 & -02 03 38.1 &  11.1 &  -8.9 & 12.7 & -1.58 & 0.76 &\\
  RXJ0254.8-0709SE& 2 54 52.43 & -07 09 24.2 &   4.7 &   5.8 & 14.5 & -0.78 & 0.59 &\\
  RXJ0255.8-0750S & 2 55 52.63 & -07 50 39.0 &  -3.4 & -15.3 & 14.0 & -1.06 & 0.81 &\\
  RXJ0309.1+0324N & 3 09 09.86 &  03 23 44.4 &  25.7 &   5.6 & 10.2 & -2.33 & 0.59 &\\
  RXJ0312.8-0414NW& 3 12 50.43 & -04 14 08.1 &  20.8 & -18.4 &  9.9 & -2.22 & 0.76 &\\
  RXJ0314.8-0406  & 3 14 50.70 & -04 05 56.4 & -14.6 &  -5.5 & 14.4 & -1.16 & 0.70 &\\
  RXJ0317.9+0231  & 3 17 59.18 &  02 30 12.9 &  11.6 &  32.9 & 11.0 & -2.26 & 0.70 &\\
  RXJ0319.3+0003  & 3 19 22.53 &  00 02 32.4 &  -0.2 &  17.6 &  9.2:& -3.24 & 0.27 &SAO 130417, (1) \\
  RXJ0324.4+0231  & 3 24 25.22 &  02 31 01.2 &   3.6 &  -7.7 & 12.4 & -1.45 & 0.65 &\\
  RXJ0329.1+0118  & 3 29 08.01 &  01 18 05.3 & -15.0 &  11.4 & 10.7 & -2.52 & 0.65 &\\
  RXJ0330.7+0306N & 3 30 43.45 &  03 05 47.9 &  12.4 &   2.6 & 10.9 & -1.86 & 0.81 &\\
  RXJ0333.0+0354  & 3 33 01.54 &  03 53 38.4 &   3.0 &  14.9 & 13.3 & -1.55 & 0.70 &\\
  RXJ0333.1+1036  & 3 33 11.62 &  10 35 56.3 &  -9.6 &   9.8 & 11.7 & -1.84 & 0.65 &\\
  RXJ0336.0+0846  & 3 36 00.32 &  08 45 35.7 &  15.4 &   0.6 & 12.4 & -1.94 & 0.76 &\\
  RXJ0338.1+1224  & 3 38 09.04 &  12 23 53.1 &  16.6 &   0.6 & 10.9 & -2.55 & 0.65 &\\
  RXJ0338.3+1020  & 3 38 18.21 &  10 20 17.1 &  21.6 &  -6.2 & 11.0 & -2.30 & 0.70 &\\
  RXJ0339.6+0624  & 3 39 40.57 &  06 24 43.1 &  13.5 & -17.3 & 11.3 & -2.38 & 0.70 &\\
  RXJ0340.3+1220  & 3 40 19.26 &  12 20 18.0 &  -1.4 &  -0.3 & 12.5 & -1.36 & 0.76 &\\
  RXJ0340.5+0639  & 3 40 31.48 &  06 39 00.3 & -15.4 &  11.8 & 13.2 & -1.79 & 0.81 &\\
  RXJ0341.2+0453  & 3 41 14.30 &  04 53 25.4 &   3.2 & -11.0 & 15.2 & -0.17 & 0.70 &\\
  RXJ0341.2+0759  & 3 41 14.36 &  07 59 33.1 &   7.1 &  -3.7 & 10.7 & -2.65 & 0.70 &\\
  RXJ0343.6+1039  & 3 43 40.51 &  10 39 14.0 &  -1.9 &  -9.3 & 10.2 & -2.50 & 0.65 &\\
  RXJ0344.8+0359  & 3 44 53.15 &  03 59 30.9 &   9.0 &   5.6 & 12.3 & -1.50 & 0.81 &\\
  RXJ0347.2+0933SW& 3 47 14.25 &  09 32 53.0 &  21.5 &  -1.4 & 13.4 & -1.24 & 0.65 &\\
  RXJ0347.9+0616  & 3 47 56.81 &  06 16 07.0 &  19.1 &  -4.8 & 11.0 & -2.40 & 0.70 &\\
  RXJ0348.2+1109  & 3 48 16.50 &  11 08 41.0 &  -4.6 &  -0.8 & 10.1:& -2.34 & 0.70 &HD 23793 B, (2) \\
  RXJ0348.5+0832  & 3 48 31.42 &  08 31 37.5 &  15.9 &  -7.5 & 10.9 & -2.04 & 0.76 &\\
  RXJ0349.4+1255N & 3 49 27.85 &  12 54 41.4 &  -6.5 &   6.1 &  9.1:& -2.83 & 0.65 &BD+12 511 \\
  RXJ0350.2+0849  & 3 50 12.91 &  08 49 34.0 &  -4.5 &  -6.1 & 12.1 & -2.09 & 0.81 &\\
  RXJ0350.6+0454  & 3 50 41.72 &  04 53 44.7 & -28.8 &  -3.4 & 15.2 & -0.82 & 0.76 &\\
  RXJ0350.6+1033  & 3 50 40.80 &  10 32 51.1 &  -9.3 &  19.9 & 10.1 & -2.70 & 0.59 &\\ 
  RXJ0351.4+0953W & 3 51 26.30 &  09 53 37.0 &   2.6 & -16.5 & 12.6 & -1.23 & 0.65 &\\
  RXJ0351.8+0413  & 3 51 49.40 &  04 13 30.7 & -16.5 & -12.3 & 13.5 & -1.51 & 0.81 &\\
  RXJ0352.4+1223  & 3 52 24.68 &  12 22 44.2 &  -4.2 &  -6.9 &  9.5:& -2.93 & 0.59 &BD+11 533 \\
\noalign {\smallskip}
\hline
\noalign {\smallskip}
\end{tabular}
\end {flushleft}
\end {table*}

\addtocounter{table}{-1}

\begin{table*}

\caption []{Continued}
\begin {flushleft}
\begin{tabular}{lrrrrrccl}
\noalign {\smallskip}
 \hline
\noalign {\smallskip}
  designation & $\alpha$ (opt.) & $\delta$ (opt.) & $\Delta \alpha (\arcsec)$ & $\Delta \delta (\arcsec)$ &
    $V$ & $\log (f_{\rm X}/f_{\rm V})$ & $P$ & remarks \\
\noalign {\smallskip}
 \hline
\noalign {\smallskip}
  RXJ0354.1+0528  & 3 54 06.58 &  05 27 23.7 &  11.3 &  20.8 & 11.9 & -2.18 & 0.70 &\\
  RXJ0354.3+0535  & 3 54 21.28 &  05 35 40.9 &  -0.5 & -14.2 & 10.1 & -2.51 & 0.54 &\\
  RXJ0354.4+1204  & 3 54 25.15 &  12 04 07.9 &  -3.3 & -14.9 & 11.2 & -2.36 & 0.70 &\\
  RXJ0354.8+1232  & 3 54 50.70 &  12 32 06.4 &   4.9 &  10.7 & 12.9 & -1.40 & 0.81 &\\
  RXJ0356.7+0943  & 3 56 45.51 &  09 43 39.4 &  -4.0 & -13.2 & 13.3 & -1.44 & 0.70 &\\
  RXJ0357.3+1258  & 3 57 21.34 &  12 58 17.3 &   3.8 &   3.6 & 10.9 & -2.16 & 0.70 &\\
  RXJ0358.1+0932  & 3 58 12.62 &  09 32 21.6 & -13.5 &  -2.1 & 12.2 & -1.42 & 0.70 &\\
  RXJ0400.0+0730  & 4 00 01.14 &  07 30 10.4 &  -0.7 &   4.0 & 12.2 & -1.86 & 0.65 &\\
  RXJ0400.1+0818N & 4 00 09.37 &  08 18 19.3 &   6.6 &   4.0 &  9.9:& -1.54 & 0.65 &BD+07 582B, (3) \\
  RXJ0400.8+1116  & 4 00 53.92 &  11 15 27.8 & -25.2 &  21.9 & 13.8 & -1.07 & 0.70 &\\
  RXJ0402.5+0552  & 4 02 35.68 &  05 51 36.0 & -18.0 &  14.5 & 10.9 & -2.08 & 0.70 &\\
  RXJ0403.5+0837  & 4 03 29.69 &  08 37 14.2 &  35.6 &   4.7 & 12.6 & -1.42 & 0.76 &\\
  RXJ0404.4+0519  & 4 04 28.48 &  05 18 43.7 &  -5.1 &   4.0 & 10.9 & -2.02 & 0.81 &\\
  RXJ0405.5+0324  & 4 05 30.24 &  03 23 50.4 &  -2.1 &   6.5 & 11.5 & -1.61 & 0.81 &\\
  RXJ0407.2+0113N & 4 07 16.49 &  01 13 14.5 &  15.2 &  -4.3 & 10.5 & -2.50 & 0.54 &\\
  RXJ0407.6+0638  & 4 07 37.26 &  06 38 45.2 &   9.5 & -23.1 & 14.8 & -0.56 & 0.54 &\\
  RXJ0408.6+1017  & 4 08 39.90 &  10 17 32.6 & -13.5 & -26.8 & 12.6 & -1.84 & 0.65 &\\
  RXJ0408.8+1028  & 4 08 49.57 &  10 27 50.0 &  -3.8 &  -4.3 &  8.9:& -2.08 & 0.70 &HD 26172 \\
  RXJ0409.8+1209  & 4 09 51.54 &  12 09 02.3 &  22.0 &  -7.6 & 12.0 & -1.81 & 0.65 &HD 286556 \\
  RXJ0410.6+0608  & 4 10 39.66 &  06 08 38.9 & -22.2 & -19.8 & 12.9 & -1.56 & 0.86 &\\
  RXJ0413.2+1028  & 4 13 14.66 &  10 28 01.1 & -10.7 & -14.7 & 14.9 & -0.56 & 0.65 &\\
  RXJ0418.6+0143  & 4 18 39.27 &  01 42 10.5 &  13.5 &  29.9 & 12.3 & -1.83 & 0.59 &\\
  RXJ0419.8+0214  & 4 19 49.92 &  02 13 43.5 & -13.8 & -11.3 & 14.6 & -0.61 & 0.76 &\\
  RXJ0419.9+0231  & 4 19 59.60 &  02 30 30.6 & -11.5 &  24.1 & 13.6 & -0.54 & 0.76 &\\
  RXJ0422.9+0141  & 4 22 54.61 &  01 41 32.1 &  22.5 & -17.7 & 12.3 & -1.63 & 0.70 &\\
  RXJ0423.5+0955  & 4 23 30.28 &  09 54 30.0 &  10.6 &  13.8 & 11.6 & -1.66 & 0.49 &\\
  RXJ0425.5+1210  & 4 25 35.30 &  12 10 00.0 &  -6.3 &  -0.5 & 10.4 & -2.40 & 0.65 &HD 286753 \\
  RXJ0426.4+0957W & 4 26 26.77 &  09 56 59.7 &  11.6 &  -3.4 & 12.0 & -1.91 & 0.81 &\\
  RXJ0427.4+1039  & 4 27 30.29 &  10 38 48.9 & -11.1 &   1.7 & 11.3 & -2.05 & 0.65 &\\
  RXJ0427.5+0616  & 4 27 32.07 &  06 15 52.1 &   0.4 &   1.8 & 10.6 & -1.80 & 0.76 &\\
  RXJ0427.8+0049  & 4 27 52.84 &  00 49 25.9 &  -5.5 &   3.3 &  9.0:& -2.92 & 0.22 &BD+00 760, (4) \\
  RXJ0429.9+0155  & 4 29 56.94 &  01 54 48.9 &   0.8 &  12.0 & 11.1 & -2.19 & 0.70 &\\
  RXJ0433.7+0522  & 4 33 46.09 &  05 22 09.1 &  18.8 & -16.6 & 15.3 & -0.71 & 0.81 &\\
  RXJ0434.3+0226  & 4 34 19.51 &  02 26 26.1 &   5.8 &   2.1 & 13.3 & -1.12 & 0.86 &\\
  RXJ0435.5+0455  & 4 35 31.56 &  04 55 32.3 &  -8.9 & -12.6 &  9.7 & -2.78 & 0.49 &\\
  RXJ0441.9+0537  & 4 41 57.64 &  05 36 34.3 &   0.3 &   7.3 & 10.2:& -1.78 & 0.70 &BD+05 706  \\
  RXJ0442.3+0118  & 4 42 18.61 &  01 17 39.9 &   8.9 &   3.5 & 11.2 & -2.03 & 0.70 &\\
  RXJ0442.5+0906  & 4 42 31.94 &  09 06 01.7 & -16.5 &  10.1 & 10.6:& -1.94 & 0.76 &BD+08 742 \\
  RXJ0442.6+1018  & 4 42 40.81 &  10 17 44.8 &  -3.6 &   3.3 &  8.2 & -2.59 & 0.65 &\\
  RXJ0442.9+0400  & 4 42 54.69 &  04 00 11.7 &   9.6 &   1.7 & 10.9 & -2.21 & 0.70 &\\
  RXJ0444.3+0941  & 4 44 20.21 &  09 41 06.8 &   1.5 &   0.3 &  8.6:& -2.60 & 0.65 &HD 287017  \\
  RXJ0444.4+0725  & 4 44 27.16 &  07 24 59.8 &  -9.3 &   8.7 & 13.7 & -1.22 & 0.65 &\\
  RXJ0444.7+0814  & 4 44 45.44 &  08 13 47.6 & -20.1 &   4.5 & 11.8 & -1.86 & 0.70 &\\
  RXJ0445.2+0729  & 4 45 13.18 &  07 29 17.7 &  -9.9 &  10.4 & 12.2 & -1.97 & 0.81 &\\
  RXJ0445.3+0914  & 4 45 23.81 &  09 13 47.8 &  -4.9 & -13.0 & 11.8 & -2.15 & 0.65 &\\
  RXJ0445.5+1207  & 4 45 36.49 &  12 07 51.1 &  -7.4 & -22.4 & 12.9 & -1.74 & 0.76 &\\
  RXJ0448.0+0738  & 4 48 00.86 &  07 37 56.9 &  -5.6 &  -7.0 & 11.2 & -1.40 & 0.70 &\\
  RXJ0450.0+0151  & 4 50 04.68 &  01 50 43.1 &  17.0 &  13.2 & 12.2 & -1.63 & 0.86 &\\
  RXJ0451.6+0619  & 4 51 41.51 &  06 19 20.6 & -12.4 & -18.1 & 12.8 & -1.73 & 0.86 &\\
  RXJ0459.9+1017  & 4 59 54.95 &  10 17 18.2 & -12.6 &   0.9 & 14.4 & -0.62 & 0.76 &\\
  RXJ0511.2+1031  & 5 11 15.98 &  10 30 36.0 &  -1.2 &   8.9 & 14.0 & -1.02 & 0.86 &\\
RXJ0511.9+1112  & 5 12 00.30 &  11 12 19.8 &  -9.5 & -14.1 & 11.4 & -2.21 & 0.65 &\\
  RXJ0512.0+1020  & 5 12 03.21 &  10 20 06.8 &  -7.5 &   5.3 & 11.3 & -1.99 & 0.76 &\\
  RXJ0513.6+0955  & 5 13 40.44 &  09 54 50.7 & -13.5 &   2.3 & 12.0 & -2.08 & 0.65 &\\
\noalign {\smallskip}
\hline
 
\end{tabular}
\end {flushleft}
\end {table*}

\addtocounter{table}{-1}

\begin{table*}
\caption []{Continued}
\begin {flushleft}
\begin{tabular}{lrrrrrccl}
\noalign {\smallskip}
 \hline
\noalign {\smallskip}
  designation & $\alpha$ (opt.) & $\delta$ (opt.) & $\Delta \alpha (\arcsec)$ & $\Delta \delta (\arcsec)$ &
    $V$ & $\log (f_{\rm X}/f_{\rm V})$ & $P$ & remarks \\
\noalign {\smallskip}
 \hline
\noalign {\smallskip}
  RXJ0515.3+1221  & 5 15 20.55 &  12 21 14.1 &  12.6 &  -1.1 & 11.6 & -1.84 & 0.81 &\\ 
  RXJ0516.3+1148  & 5 16 21.52 &  11 47 47.3 &   5.3 &  -3.2 & 12.4 & -1.53 & 0.86 &\\
  RXJ0523.0+0934  & 5 23 04.65 &  09 34 01.0 & -25.2 &   5.0 & 10.4 & -2.64 & 0.76 &\\
  RXJ0523.5+1005  & 5 23 33.72 &  10 04 29.8 &  -1.5 & -11.1 & 11.2 & -1.81 & 0.76 &\\
  RXJ0523.9+1101  & 5 23 57.04 &  11 00 57.5 & -26.4 &  15.9 & 10.8 & -2.62 & 0.70 &\\
  RXJ0525.7+1205SE& 5 25 47.21 &  12 04 10.2 &   0.6 &  36.3 & 14.1 & -1.17 & 0.65 &\\
  RXJ0528.4+1213  & 5 28 25.74 &  12 12 36.1 &  -4.9 &   5.9 & 11.5 & -1.77 & 0.81 &\\
  RXJ0528.5+1219  & 5 28 35.19 &  12 19 04.0 &   1.8 &  10.1 & 12.7 & -1.77 & 0.70 &\\
  RXJ0528.9+1046  & 5 28 58.50 &  10 45 37.9 &  -7.9 &  -0.1 & 12.7 & -1.71 & 0.76 &\\
  RXJ0529.3+1210  & 5 29 18.97 &  12 09 29.7 &  -4.6 &   3.6 & 12.9 & -1.36 & 0.54 &\\
  RXJ0530.9+1227  & 5 30 57.23 &  12 27 26.8 &  27.3 &   0.4 & 10.7 & -2.19 & 0.70 &\\
  RXJ0531.8+1218  & 5 31 47.77 &  12 18 08.1 &  16.4 &  12.8 & 12.1 & -2.06 & 0.81 &\\
\noalign{\smallskip}
 \hline
\noalign {\smallskip}
\end {tabular}

\smallskip

\noindent Remarks:
(1) Not listed in the GSC;
(2) HD 23793~B (F5V) and the early-type  star HR~1174 (B3) are an optical
pair with $9\arcsec$ separation (Lindroos 1986); the later type star is more likely to be the true
optical counterpart of the X-ray source (e.g. Schmitt et al. 1993);
(3) Companion N is an unresolved binary itself;
(4) Not listed in the GSC.

\end {flushleft}
\end{table*}

In Table 4, we present the results of our spectroscopic observations.
We list all the counterparts observed (with names from Tables 2 and 3),
identify the telescope, i.e. the resolution and wavelength range used,
give equivalent widths of the \ha\ and \li\ lines, and list the
spectral types. In the lithium column, ``no'' means that no lithium
line has been detected above the noise. Where possible, upper limits
have been estimated (but not reported here), resulting always lower
than 0.1~\AA. The typical error  in the Li equivalent width is $\pm
20$\,m\AA, mainly due to   uncertainties in the location of the
continuum.

\section{Results}

Our  aim is to search for low-mass PMS stars among the RASS source
counterparts observed in this work, by studying their  \li\ doublet.
The presence of a strong \li\ absorption is the most evident indicator
of the PMS nature. Typical equivalent widths range from 0.7~\AA,
observed in cTTS (Magazz\`u \etl\ 1992) to 0.1~\AA, observed in wTTS
(Mart\'\i n \etl\ 1994), while equivalent widths lower than about
0.3~\AA\ have been measured in Pleiades objects (Soderblom \etl\ 1993,
Garc\'\i a L\'opez \etl\ 1994).

\begin {figure}
\psfig {file=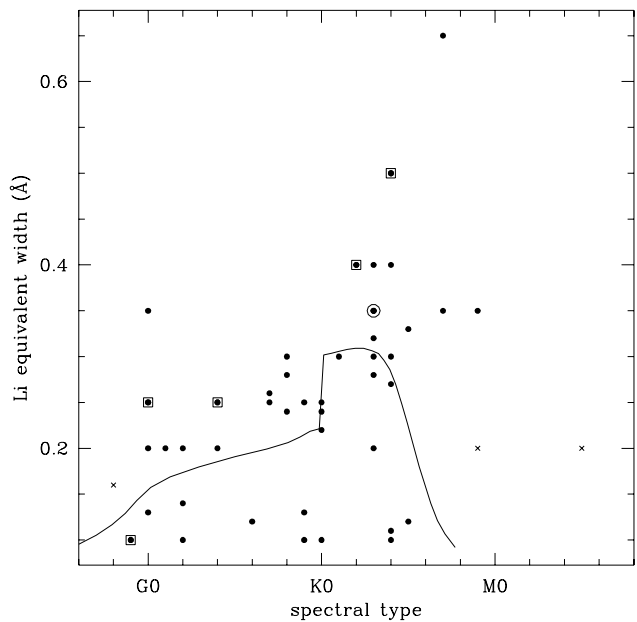,height=8.8 cm}
\caption[]{Equivalent widths of the \li\ doublet vs.\ spectral type for
the stars in our sample. The continuous line has been adapted from
Fig.~2d in Soderblom \etl\ (1993) and represents the upper envelope of
Li equivalent width for Pleiades stars. Squares (circles) indicate two
(three) points in the same position. Crosses are doubtful points (see
text)} 
\end {figure}

\begin{table*}[t]
\caption []{Other optical counterparts observed.  Listed are optical
counterparts other than those in Table 2, but which have also been
observed by us.  We list designation (as in Table 1, but with
indications regarding the particular counterpart listed here),
approximate optical position (for J\,2000.0), a remark regarding its
position relative to the counterpart listed in Table 2, approximate V
magnitude, and remarks.  The source for optical positions, V
magnitudes, and remarks are either GSC, Simbad, or visual inspection of
plates}

\begin {flushleft}
\begin{tabular}{lrrlll} 
\noalign {\smallskip}
 \hline
\noalign {\smallskip}
  designation & $\alpha$ (opt.) & $\delta$ (opt.) & rel. pos. & $V$ & remarks \\ \noalign {\smallskip}
\hline
\noalign {\smallskip}
    RXJ0210.4$-$1308SW & 2 10 26   & -13 07 56 & 2$\arcsec$ southwest of NE  & 12.5     & \\
    RXJ0219.4$-$1321A  & 2 20 29   & -13 20 25 & 60$\arcsec$ northeast of B  & 13       & \\
    RXJ0219.4$-$1321C  & 2 19 25   & -13 21 47 & 40$\arcsec$ south of B      & 15.5     & \\
    RXJ0223.3$-$1615SW & 2 23 25   & -16 14 31 & 10$\arcsec$ southwest of NE & 13.5     & \\
    RXJ0254.8$-$0709NW & 2 54 53   & -07 09 20 & 6$\arcsec$ northwest of SE  & 15.5     & \\
    RXJ0255.8$-$0750N  & 2 55 52   & -07 50 29 & 10$\arcsec$ north of S      & 15.5     & \\
    RXJ0309.1$+$0324S  & 3 09 10   &  03 23 42 & 2$\arcsec$ south of N       & 11       & \\
    RXJ0312.8$-$0414SE & 3 12 51   & -04 14 19 & 14$\arcsec$ southeast of NW & 10.5     & \\
    RXJ0330.7$+$0306S  & 3 30 43   &  03 05 18 & 30$\arcsec$ south of N      & 15       & \\
    RXJ0347.2$+$0933NE & 3 47 17   &  09 33 08 & 40$\arcsec$ northeast of SW & 12       & \\
    RXJ0349.4$+$1255S  & 3 49 28   &  12 54 28 & 14$\arcsec$ south of N      & 10       & BD+12 511B \\
    RXJ0351.4$+$0953E  & 3 51 28   &  09 53 34 & 25$\arcsec$ east of W       & 13.7     & \\
    RXJ0400.1$+$0818S  & 4 00 09   &  08 18 15 & 4$\arcsec$ south of N       & 10       & BD+07 582 \\
    RXJ0407.2$+$0113S  & 4 07 16   &  01 13 12 & 2$\arcsec$ south of N       & 12       & \\
    RXJ0426.4$+$0957E  & 4 27 30   &  09 57 00 & 45$\arcsec$ east of W       & 11.5     & also in GSC \\
    RXJ0525.7$+$1205NW & 5 25 44   &  12 04 30 & 45$\arcsec$ northwest of SE & 14.5     & \\ 
\noalign {\smallskip}
\hline

\end{tabular}
\end {flushleft}

\end{table*}

In Fig.~3 we show, for our objects in which Li has been detected, the
\li\ equivalent width vs.\ the spectral type. We plot also the upper
envelope of \li\ equivalent widths for stars in the Pleiades (adapted
from Fig.~2d in Soderblom \etl\ 1993).

In this paper we classify as low-mass PMS stars those objects, later
than F7, satisfying the following criterion:
\begin {quote} 
The \li\  equivalent width must be greater than in Pleiades objects of
the same spectral type.
\end {quote}
In other words, a star is an PMS if it can be located  above the
continuous line in Fig.~3. Stars in which lithium has been detected,
but below the Pleiades upper envelope, will be classified as possible
PMS stars (``PMS?'' in Table~4). Objects with no lithium will be
classified as dKe or dMe stars, depending on their spectral type, if
their \ha\ is in emission or strongly filled in. The remaining objects
will be classified as ``non-PMS''\@. We are aware that this
classification is somewhat arbitrary, for example among the non-PMS we
could find some post-TTS, which in some cases show the Li resonance
doublet weaker than 0.1~\AA\ (Mart\'\i n \etl\ 1992). However, we
believe this  is a conservative classification, which makes us quite
confident that objects classified as ``PMS'' are really PMS stars. In
order to get accurate ages of our objects, we are planning further
observations, aimed  to locate our PMS candidates in the H-R diagram
and measure their lithium abundance.

Note that we do not impose any restriction on \ha\ for PMS stars. In
fact, TTS show a wide range of \ha\ equivalent widths, from strong
emission in cTTS to practically no emission in post-TTS. However, from
Table~4 we can see that most of the objects classified as PMS or PMS?
show weak \ha\ emission or \ha\ absorption shallower than in main
sequence objects, as expected for wTTS.

Although the objects RXJ0255.8$-$0750N, RXJ0333.0 $+$0354, and
RXJ0422.9$+$0141 in Fig.~3 are located above the Pleiades upper
envelope, in this paper we classify them as PMS?\@. For the first two
objects this is due to the quite high level of noise, which makes
doubtful the detection of the lithium line.  For RXJ0422.9$+$0141 we
note that this object is a spectroscopic binary  (see remark~7 in
Table~4) and the determination of the spectral type of its components
is rather uncertain.

\begin{figure*}
\centerline{\psfig {file=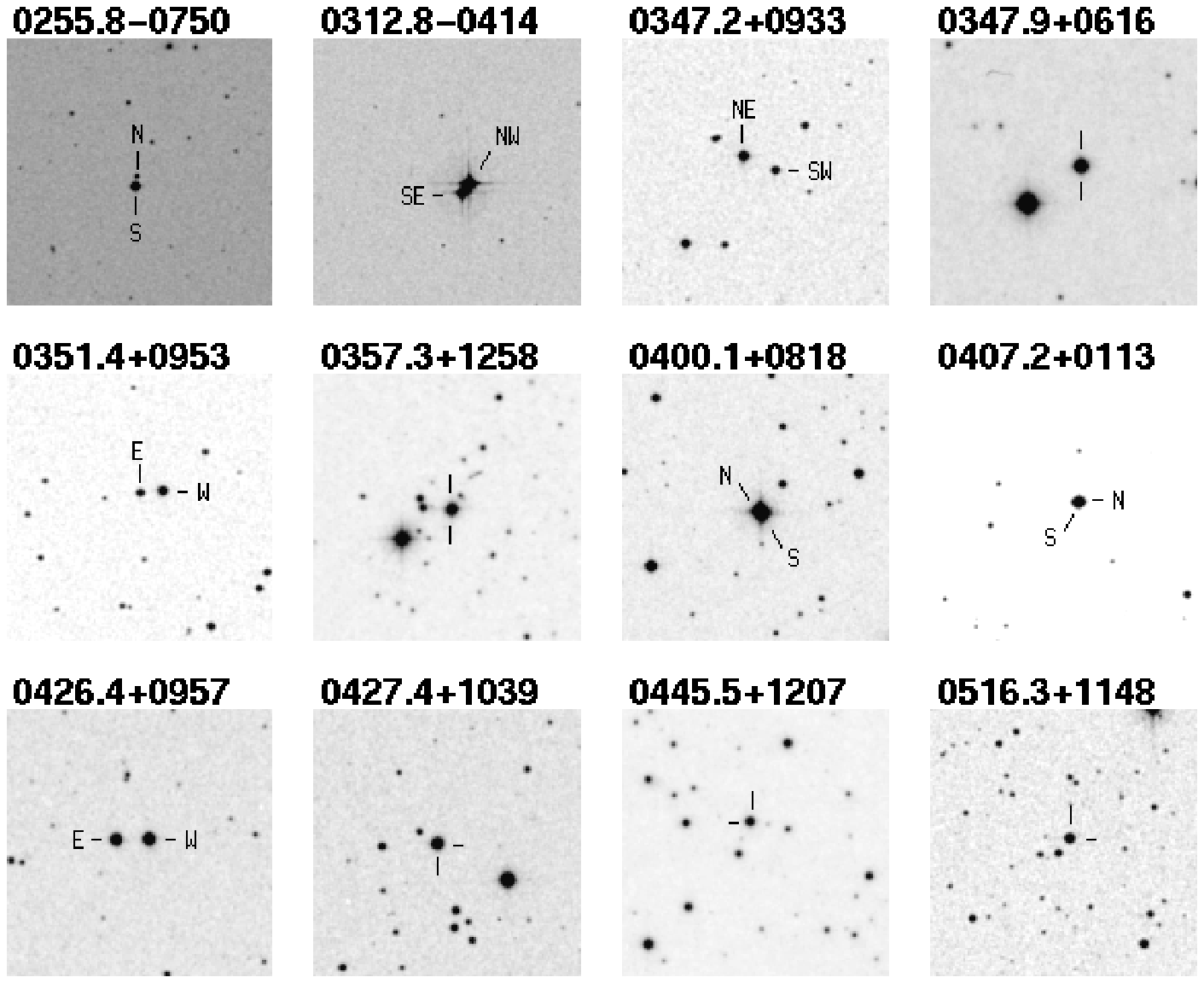,bbllx=70pt,bblly=371pt,bburx=500pt,bbury=754pt,clip=,height=12cm}}
\caption[]{Some finding charts (5\arcmin $\times$ 5\arcmin). North is up, East on the left. The other PMS and PMS? stars can be easily identified (see text)}
\end{figure*}

In total, there are 115 RASS sources selected for optical follow-up
observations. As some RASS sources have several potential optical
counterparts, we have performed spectroscopy for a total of 131 stars,
all listed in Table 4. In Fig.~4 we show the finding charts for some
object whose identification may be difficult; all the other new PMS can
be easily identified, as they are GSC stars with no bright stars in the
vicinity. In case of need, the charts for any object observed here are
available on request.

There are 30 stars in our sample which can be classified as PMS stars,
according to our criterion. These objects are counterparts to 28 RASS
sources, i.e.\ for two RASS sources we have found two new PMS stars
each (with separations of $2\arcsec$ and $14\arcsec$).  In addition,
there are 19 RASS counterparts classified as PMS? as well as 17 and 5
RASS counterparts classified as dKe and  dMe, respectively. In several
spectra, we clearly see double lines indicative of close binaries
(identified as spectroscopic binaries -- ``SB'' -- in Table~4). Namely,
we find seven (one) SB among the stars classified as dKe (dMe) stars,
three  among the PMS? stars, and four  among the non-PMS stars.

The spectra of all 30  stars classified in Table~4 as new PMS stars are
shown in Fig.~5. These and all the other spectra are available from the
authors.  In some cases one can see quite broad lines indicative of
either close binaries and/or fast rotation. As some relatively fast
rotators are expected to be present among TTS,  we do not classify such
PMS stars as SB\@.

As  can be seen from Table~4, three of the 15 stars classified as wTTS
in Neuh\"auser \etl\ (1995c) may not be real young PMS stars according
to our   stricter criterion; here they are classified as dKe stars.
Seven other stars classified as wTTS by Neuh\"auser \etl\  are now
classified as PMS? stars, while the other five wTTS  are confirmed to
be PMS stars also in this paper.

We note that one of the stars studied here (HD 23793 B) is the
late-type secondary in the so-called Lindroos sample (Lindroos 1986), a
set of double stars with the primary being an early-type star and the
secondary being late-type. It has been suggested that many of the
secondaries in this sample may be post-TTS (Lindroos 1986).  However,
both Pallavicini \etl\ (1992) and Mart\'\i n \etl\ (1992) consistently
found that only about one third of them may be genuine post-TTS. As far
as HD 23793 B is concerned (spectral type F5), both Pallavicini
\etl\ (1992) and ourselves find \ha~absorption and very weak \li .
Pallavicini \etl\ (1992) did not detect Ca H and K emission and
classified this Lindroos pair as an optical system.  According to our
criterion, we classify this star as non-PMS as well.
 
\section{Discussion}

From Fig.~1 we see that many of our newly discovered low-mass PMS stars
are located up to several tens of degrees away from regions of ongoing
star formation. In fact, the area of on-going star formation in Taurus,
as defined by the CO contours (Ungerechts \& Thaddeus 1987) and also
the area populated by the TTS  known before ROSAT or newly discovered
by Wichmann \etl\ (1996), is  at $\delta > 14\degr$\@. If the distance
of our objects is the same as the Taurus-Auriga clouds (140~pc), these
objects will lie up to several tens of parsecs from the clouds. In any
case, they are located far from any known star forming region. Despite
their location, the lithium equivalent width in most of these stars is
indicative of young ages, just typical for wTTS.

\begin {table*}

\caption []{Results of medium-resolution spectroscopy of all potential
counterparts  to RASS X-ray sources. Listed are source designation, the
telescope used to observe the star (INT for the Isaac Newton Telescope,
ESO for the ESO\,1.52m telescope), H$\alpha$ equivalent width (negative
when in emission, `f' indicates that $H \alpha$ emission is filling in
the absorption line, `abs' for absorption without available equivalent
width), lithium $6708$\AA\ absorption equivalent width (colons indicate
large uncertainty), spectral type (always luminosity class V unless
otherwise noted), nature of object (see text), and remarks (e.g. `SB'
for spectroscopic binaries or `(N95c)' if mentioned in Neuh\"auser
\etl\ 1995c)}

\begin {flushleft}
\begin{tabular}{lcrrlll}
\noalign{\smallskip}
 \hline
\noalign{\smallskip}
   designation & tel. & $W_{\lambda}(H \alpha)$ & $W_{\lambda}$(Li) 
   & SpType & nature & remarks \\ 
~ & & (\AA) & (\AA) & & & \\ 
\noalign{\smallskip}
\hline
\noalign{\smallskip}
  RXJ0207.0$-$1407  & INT & 3.3      & no   & F8   & non-PMS & \\
  RXJ0209.1$-$1536  & INT & -1.2     & no   & K4   & dKe     & \\
  RXJ0210.4$-$1308SW& INT & -0.12    & no   & K3   & dKe     &  2$\arcsec$ sep. \\
  RXJ0210.4$-$1308NE& INT & 3.5      & no   & G5   & non-PMS &  2$\arcsec$ sep. \\
  RXJ0212.3$-$1330  & INT & 1.2      & no   & K1   & non-PMS & \\
  RXJ0215.0$-$1402  & INT & 1.15     & no   & K4   & non-PMS & maybe K3 III \\
  RXJ0218.6$-$1004  & INT & 1.6      & no   & G8   & non-PMS & SB \\
  RXJ0219.4$-$1321C & ESO & abs      & no   & G0   & non-PMS & \\
  RXJ0219.4$-$1321B & ESO & 0.60 f   & no   & M0   & dMe     & \\
  RXJ0219.4$-$1321A & ESO & -4.30    & no   & M3   & dMe     & \\
  RXJ0219.7$-$1026  & INT & -0.25    & 0.1:   & K4   & PMS?    & (N95c),  (1)  \\
  RXJ0220.4$-$1305  & INT & 5.7      & no   & F0?  & non-PMS & \\
  RXJ0223.3$-$1615SW& ESO & -0.8     & no   & K7   & dKe     &  $10\arcsec$ sep.\\
  RXJ0223.3$-$1615NE& ESO & 2.1      & no   & G7   & non-PMS & $10\arcsec$ sep. \\
  RXJ0229.5$-$1224  & ESO & 2.6      & 0.28 & G8   & PMS     & HD~15526\\
  RXJ0237.3$-$0527  & INT & -0.30    & no   & K5   & dKe     & (N95c) \\
  RXJ0239.1$-$1028  & INT & -0.35    & no   &K7-M0 & dKe    & SB \\
  RXJ0243.9$-$0850  & ESO & 0.50 f   & no   & M2   & dMe     & \\
  RXJ0248.3$-$1117  & INT & 2.1      & no   & G7   & non-PMS & \\
  RXJ0251.8$-$0203  & INT & -0.60    & no   & K6   & dKe     & (N95c) \\
  RXJ0254.8$-$0709NW& INT & -4.2     & no   & M5   & dMe     &  3$\arcsec$ sep. \\
  RXJ0254.8$-$0709SE& INT & -2.2     & no   & M3   & dMe     &  3$\arcsec$ sep. \\
  RXJ0255.8$-$0750N & INT & -0.7     & 0.2: & M5   & PMS?   &  $10\arcsec$ sep. \\
  RXJ0255.8$-$0750S & INT & -0.2     & no   &K7-M0 & dKe     & $10\arcsec$ sep., SB   \\
  RXJ0309.1$+$0324N & INT & 3.6      & no   & F7   & non-PMS &  2$\arcsec$ sep. \\
  RXJ0309.1$+$0324S & INT & 1.7      & no   & G6   & non-PMS &  2$\arcsec$ sep. \\
  RXJ0312.8$-$0414NW& ESO & 3.5      & 0.2  & G0   & PMS     & $14\arcsec$ sep.\\
  RXJ0312.8$-$0414SE& ESO & 2.5      & 0.3  & G8   & PMS     & $14\arcsec$ sep. \\
  RXJ0314.8$-$0406  & INT & 5.5      & no   &late A& non-PMS & \\
  RXJ0317.9$+$0231  & INT & 1.4      & no   & G6   & non-PMS & \\
  RXJ0319.3$+$0003  & ESO & 4.4      & no   & G5   & non-PMS & SAO 130417 \\
  RXJ0324.4$+$0231  & INT & -0.40    & 0.33 & K5   & PMS     & (N95c) \\
  RXJ0329.1$+$0118  & INT & 4.0      & 0.13 & G0   & PMS?     & (N95c) \\
  RXJ0330.7$+$0306N & INT & 1.1 f    & no   & K5   & dKe     & SB, (2) \\
  RXJ0330.7$+$0306S & INT & 1.0      & no   & K7   & non-PMS & \\
  RXJ0333.0$+$0354  & INT & -2.4     & 0.2:  &K7-M0 & PMS?   & (N95c)  \\
  RXJ0333.1$+$1036  & INT & -0.8     & 0.32 & K3   & PMS     & (N95c) \\
  RXJ0336.0$+$0846  & INT & -0.1     & no   & M3   & dMe     & \\
  RXJ0338.1$+$1224  & ESO & 1.5      & no   & K0   & non-PMS & \\
  RXJ0338.3$+$1020  & INT & 2.0      & 0.25 & G9   & PMS     & \\
  RXJ0339.6$+$0624  & INT & -0.1     & 0.13  & G9   & PMS?   & (N95c)  \\
  RXJ0340.3$+$1220  & INT & -1.1     & no   & K5   & dKe     & \\
  RXJ0340.5$+$0639  & ESO & 1.6      & no   & K2   & non-PMS & (3) \\
  RXJ0341.2$+$0453  & INT & 3.2      & no   & G9   & non-PMS & \\
  RXJ0341.2$+$0759  & ESO & 3.2      & no   & K0   & non-PMS & \\
  RXJ0343.6$+$1039  & INT & 1.5      & 0.1  & K0   & PMS?   & SB,  (4) \\
  RXJ0344.8$+$0359  & INT & 0.3 f    & 0.30 & K3   & PMS?     & (N95c) \\
  RXJ0347.2$+$0933SW& ESO & -0.40    & 0.4  & K4   & PMS     & \\
  RXJ0347.2$+$0933NE& ESO & 2.00     & 0.1  & G9   & PMS?   &  \\
  RXJ0347.9$+$0616  & INT & 2.6      & 0.2  & G2   & PMS     & \\
\noalign{\smallskip}
\hline
\end{tabular}
\end {flushleft}
\end {table*}

\addtocounter{table}{-1}

\begin{table*}

\caption []{Continued}

\begin {flushleft}
\begin{tabular}{lcrrlll}
\noalign{\smallskip}
 \hline
\noalign{\smallskip}
   designation & tel. & $W_{\lambda}(H \alpha)$ & $W_{\lambda}$(Li) 
   & SpType & nature & remarks \\ 
~ & & (\AA) & (\AA) & & & \\ 
\noalign{\smallskip}
\hline
\noalign{\smallskip}
   RXJ0348.2$+$1109  & ESO & 5.6      & no   & F5   & non-PMS & HD~23793B  \\
 RXJ0348.5$+$0832  & INT & -0.1     & 0.26 & G7   & PMS     & (5) \\ 
  RXJ0349.4$+$1255N & INT & 2.5      & no   & G0   & non-PMS & BD+12 511, $14\arcsec$ sep. \\  RXJ0349.4$+$1255S        & INT & 2.0      & no   & G7   & non-PMS & BD+12 511B, $14\arcsec$ sep.  \\
  RXJ0350.2$+$0849  & INT & -0.2     & no   & K5   & dKe     & SB \\
  RXJ0350.6$+$0454  & INT & -0.15    & no   & K7   & dKe     & SB \\
  RXJ0350.6$+$1033  & ESO & 2.4      & no   &K0 III& non-PMS & \\
  RXJ0351.4$+$0953W & INT & 0.5 f    & 0.3  & K1   & PMS?     & \\
  RXJ0351.4$+$0953E & INT & 2.0      & no   & F0   & non-PMS & maybe F0 III \\
  RXJ0351.8$+$0413  & ESO & 2.00     & 0.12 & G6   & PMS? & (3), (6) \\ 
  RXJ0352.4$+$1223  & ESO & 3.7      & 0.10 & G2   & PMS?     & BD+11 533\\
  RXJ0354.1$+$0528  & ESO & 2.8      & 0.24 & G8   & PMS     & (3) \\
  RXJ0354.3$+$0535  & INT & 3.5     & 0.2  & G1   & PMS     & (N95c) \\
  RXJ0354.4$+$1204  & ESO & 3.9      & no   & G5   & non-PMS & \\
  RXJ0354.8$+$1232  & INT & -1.7     & no   & K7   & dKe     & \\
  RXJ0356.7$+$0943  & INT & -2.7     & no   & M3   & dMe     & \\
  RXJ0357.3$+$1258  & INT & 1.8      & 0.25 & G0:  & PMS     &  \\
  RXJ0358.1$+$0932  & INT & -0.10 f  & 0.2  & K3   & PMS?     & (N95c) \\ 
  RXJ0400.0$+$0730  & INT & 1.3      & no   & G3   & non-PMS & \\
  RXJ0400.1$+$0818S  & ESO & 2.5     & 0.24 & K0   & PMS?     & BD+07 582, 4$\arcsec$ sep. \\
  RXJ0400.1$+$0818N  & ESO & -0.05    & 0.40 & K2   & PMS     & BD+07 582B, 4$\arcsec$ sep. \\
  RXJ0400.8$+$1116  & INT & 1.8      & no   & K0   & non-PMS & \\
  RXJ0402.5$+$0552  & INT & 1.7      & no   & G4   & non-PMS & \\
  RXJ0403.5$+$0837  & INT & 1.3      & no   & K0   & non-PMS & \\
  RXJ0404.4$+$0519  & INT & 1.0 f    & 0.25 & K0   & PMS?     & \\
  RXJ0405.5$+$0324  & INT & -0.4     & no   & K4   & dKe     & \\
  RXJ0407.2$+$0113N & INT & 3.3      & 0.2  & G4   & PMS     & 2$\arcsec$ sep. \\
  RXJ0407.2$+$0113S & INT & 0.5 f    & 0.35 & K3   & PMS     & 2$\arcsec$ sep. \\
  RXJ0407.6$+$0638  & INT & 2.0      & no   & G0   & non-PMS & \\
  RXJ0408.6$+$1017  & INT & 1.8      & no   & G7   & non-PMS & \\
  RXJ0408.8$+$1028  & ESO & 3.2      & no   & G5   & non-PMS & HD~26172\\
  RXJ0409.8$+$1209  & INT & 3.3      & 0.10 & F9   & PMS?   & HD~286556 \\
  RXJ0410.6$+$0608  & INT & -0.05    & 0.11 & K4   & PMS?   & SB? \\
  RXJ0413.2$+$1028  & INT & 3.0      & no   & G0   & non-PMS & \\
  RXJ0418.6$+$0143  & INT & 0.4 f    & no   & K4   & dKe     & \\
  RXJ0419.8$+$0214  & INT & 4.5      & no   & F5   & non-PMS & \\
  RXJ0419.9$+$0231  & INT & 3.6      & no   & F9   & non-PMS & \\
  RXJ0422.9$+$0141  & INT & 1.5      & yes  & F8   & PMS?   & SB, (7) \\
  RXJ0423.5$+$0955  & INT & -0.1     & 0.27 & K4   & PMS?     & \\
  RXJ0425.5$+$1210  & INT & 4.2      & 0.1  & F9   & PMS?     & HD~286753\\
  RXJ0426.4$+$0957W & INT & 7.5      & no   &late A& non-PMS & \\
  RXJ0426.4$+$0957E & INT & 2.9      & 0.14 & G2   & PMS?     & \\
  RXJ0427.4$+$1039  & INT & 1.3 f    & 0.35 & G0:  & PMS     &  \\
  RXJ0427.5$+$0616  & INT & 1.5      & 0.25 & G4   & PMS     & \\
  RXJ0427.8$+$0049  & INT & 3.2      & no   & G3   & non-PMS & BD+00 760, SB \\
  RXJ0429.9$+$0155  & INT & 1.5      & no   & K3   & non-PMS & maybe K3 III \\
  RXJ0433.7$+$0522  & INT & 4.0      & no   & F8   & non-PMS & \\
  RXJ0434.3$+$0226  & INT & -0.4     & 0.3  & K4   & PMS     & \\ 
  RXJ0435.5$+$0455  & INT & 1.2      & no   &K3 III& non-PMS & \\
  RXJ0441.9$+$0537  & INT & abs      & no   & G5   & non-PMS & BD+05 706, (8) \\
  RXJ0442.3$+$0118  & INT & -1.0     & no   & K2   & dKe     & \\
  RXJ0442.5$+$0906  & INT & 1.4      & 0.25 & G7   & PMS     & BD+08 742 \\
  RXJ0442.6$+$1018  & INT & 1.2      & no   & K3   & non-PMS & maybe K3 III \\
  RXJ0442.9$+$0400  & INT & 1.1      & 0.22  & K0   & PMS?     & \\
\noalign{\smallskip}
 \hline
\end{tabular}
\end {flushleft}
\end {table*}

\addtocounter{table}{-1}

\begin{table*}

\caption []{Continued}

\begin {flushleft}
\begin{tabular}{lcrrlll}
\noalign{\smallskip}
 \hline
\noalign{\smallskip}
   designation & tel. & $W_{\lambda}(H \alpha)$ & $W_{\lambda}$(Li) 
   & SpType & nature & remarks \\ 
~ & & (\AA) & (\AA) & & & \\
\noalign{\smallskip}
 \hline
\noalign{\smallskip}
  RXJ0444.3$+$0941  & INT & 3.2      & no   & F9   & non-PMS & HD~287017, SB \\
  RXJ0444.4$+$0725  & INT & -0.3     & 0.12  & K5   & PMS?   &   \\
  RXJ0444.7$+$0814  & INT & -0.80    & 0.28  & K3   & PMS?     & (N95c) \\
 RXJ0445.2$+$0729  & INT & 2.6      & 0.25  & G0   & PMS     & \\
  RXJ0445.3$+$0914  & INT & 3.8      & no   & G0   & non-PMS & \\
  RXJ0445.5$+$1207  & INT & -2.0     & 0.35 & K7   & PMS     & \\
  RXJ0448.0$+$0738  & INT & -0.1 f   & no   & K1   & dKe     & (N95c) \\
  RXJ0450.0$+$0151  & INT & 0.5 f    & 0.35 & K3   & PMS     & \\
  RXJ0451.6$+$0619  & INT & 1.3 f    & no   & K2   & dKe     & SB? \\
  RXJ0459.9$+$1017  & INT & 3.0      & no   & F5   & non-PMS & SB \\ 
  RXJ0511.2$+$1031  & INT & -2.8     & 0.65 & K7   & PMS     & \\
  RXJ0511.9$+$1112  & INT & 1.2      & 0.25 & G4   & PMS     & \\
  RXJ0512.0$+$1020  & INT & -0.1     & 0.4  & K2   & PMS     & \\
  RXJ0513.6$+$0955  & INT &  1.4     & no   & G6   & non-PMS & \\
  RXJ0515.3$+$1221  & INT & 1.2 f    & no   & K0   & dKe     & SB, (9) \\
  RXJ0516.3$+$1148  & INT & 0.1 f    & 0.5  & K4   & PMS     & (N95c) \\
  RXJ0523.0$+$0934  & INT & 4.2      & no   & F8   & non-PMS & \\ 
  RXJ0523.5$+$1005  & INT & -0.5     & no   & K3   & dKe     & SB, (9) \\ 
  RXJ0523.9$+$1101  & INT & 3.7      & no   & G0   & non-PMS & \\
  RXJ0525.7$+$1205NW& INT & 1.4      & no   & G8   & non-PMS & maybe G8 III \\	
  RXJ0525.7$+$1205SE& INT & 1.4      & no   & G8   & non-PMS & maybe G8 III \\
  RXJ0528.4$+$1213  & INT & 2.5      & no   & K2   & non-PMS & \\
  RXJ0528.5$+$1219  & INT & 0.7 f    & 0.35 & K3   & PMS     & \\
  RXJ0528.9$+$1046  & INT & 0.1 f    & 0.4  & K3   & PMS     & \\
  RXJ0529.3$+$1210  & INT & -2.0     & 0.35 &K7-M0 & PMS     & \\
  RXJ0530.9$+$1227  & INT & 1.3      & no   & K0   & non-PMS & \\
  RXJ0531.8$+$1218  & INT & -0.74    & 0.5  & K4   & PMS     & (N95c) \\ 
\noalign{\smallskip}
\hline
\end {tabular}
\end {flushleft}
\smallskip

\noindent Remarks:  
(1) Spectrum blue-shifted by $2$\AA;
(2) Star A itself is SB, with the secondary 
having almost the same spectral type as the primary;
(3) Also observed at INT;
(4) Spectrum blue-shifted by $1$\AA, the $H \alpha$ line shows a P~Cyg profile;
(5) $H \alpha$ line shows an inverse P~Cyg profile;
(6) Spectrum red-shifted by $2$\AA;
(7) SB: the primary is F8  with $H \alpha$ in
absorption and $W_{\lambda}$(Ca)$~<~W_{\lambda}$(Li)$~=~0.16$\AA, 
the secondary has $W_{\lambda}$(Ca)$~>~W_{\lambda}$(Li)$~=~0.13$\AA, 
thus the system may by a close PMS binary; 
$H \alpha$ line shows a P~Cyg profile;
(8) Very noisy spectrum;
(9) Secondary seems to have a slightly earlier spectral type.

 \end{table*}

Radial velocities for some of them have been presented in Neuh\"auser
\etl\ (1995c) and indicate that about half of their 15 objects are
kinematic members of the Taurus-Auriga T~association.  A complete
analysis of the kinematic status of all stars studied here will be
given in Neuh\"auser \etl\ (1997) together with radial velocities (for
almost all stars studied here) and proper motions for several stars
identified here as new PMS stars.

In any SFR studied the RASS has revealed hundreds of new wTTS, which
have been discovered even outside regions of ongoing star formation.
However, only few cTTS have been discovered, either by EO or ROSAT.  As
the RASS is flux-limited and ROSAT pointed observations are spatially
biased towards ``interesting'' regions, many wTTS have not been
discovered yet. We find 30 new PMS stars among 115 previously
unidentified sources (i.e.\ 26\%) selected by hardness ratios and V
magnitude of the possible counterpart. Although our investigation has
been carried out outside molecular gas regions, this percentage of new
PMS stars is within the range expected in Neuh\"auser \etl\ (1995a),
who predict at least 286 new wTTS  among 1143 unidentified pre-selected
RASS sources (i.e.\ at least 25\%).  As far as the wTTS/cTTS ratio is
concerned, we note that in the area studied in this paper no cTTS are
known. In the complete area studied by Neuh\"auser \etl\ (1995a) this
ratio is about 8:1 or larger, while, considering only the dark cloud
cores, the ratio is about 1:1 (see Neuh\"auser \etl\ 1995a for a
discussion).

\begin{figure*}
\psfig {file=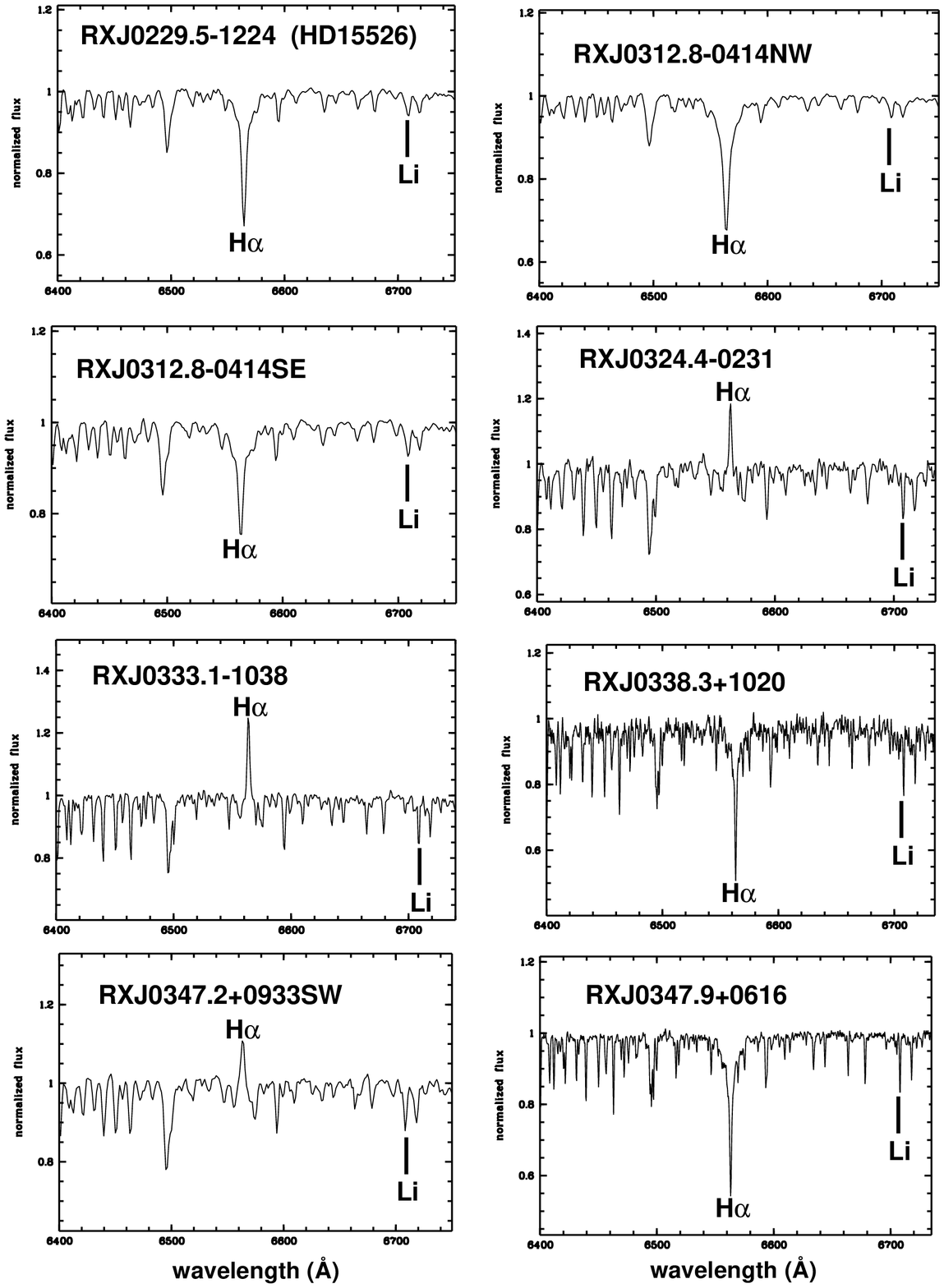,height=22.5 cm}
\caption[]{Spectra   of our  newly discovered PMS stars}
 \end{figure*}

\addtocounter{figure}{-1}

\begin{figure*}
\psfig {file=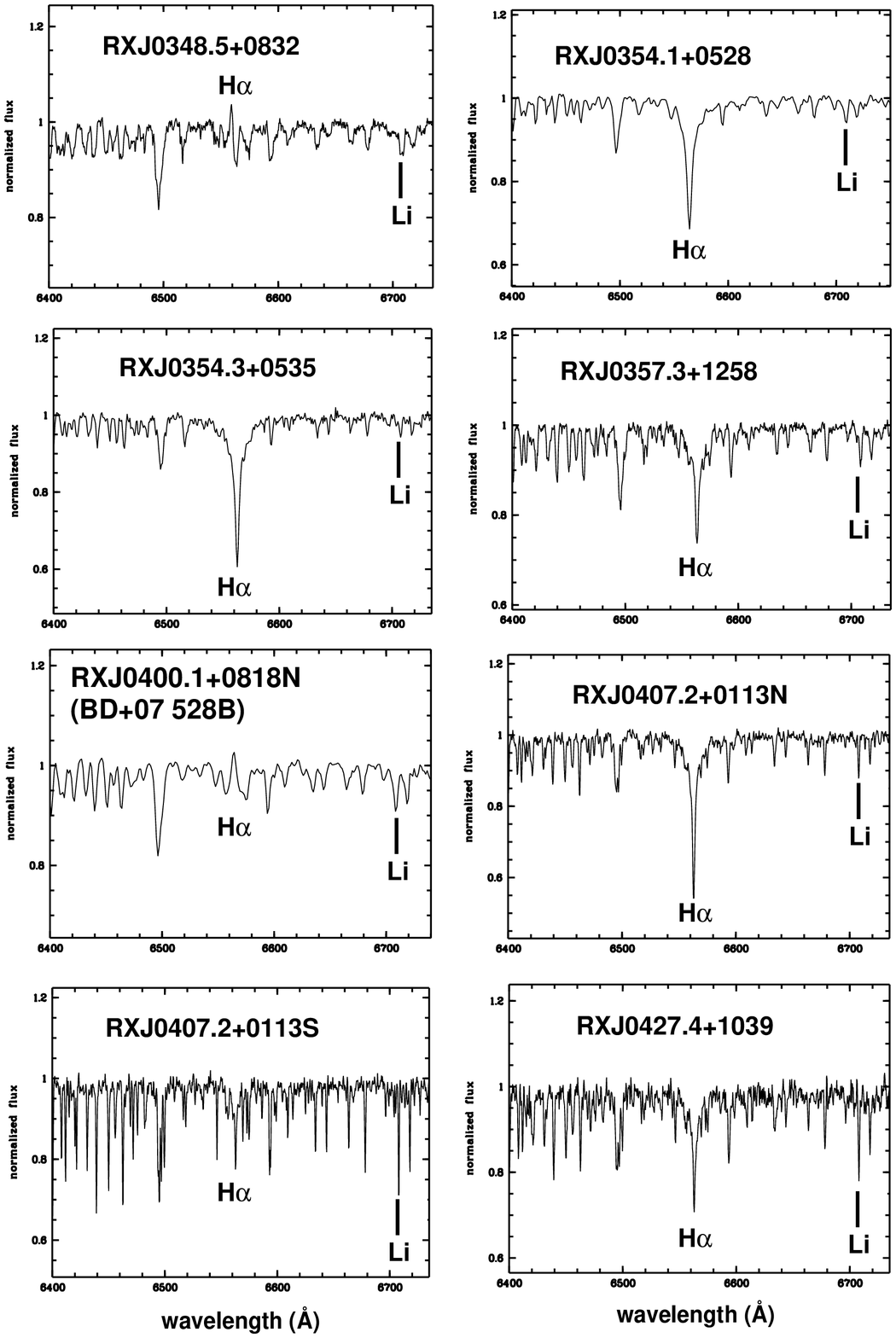,height=22.5 cm}
\caption[]{Continued}
 \end{figure*}

\addtocounter{figure}{-1}

\begin{figure*}
\psfig {file=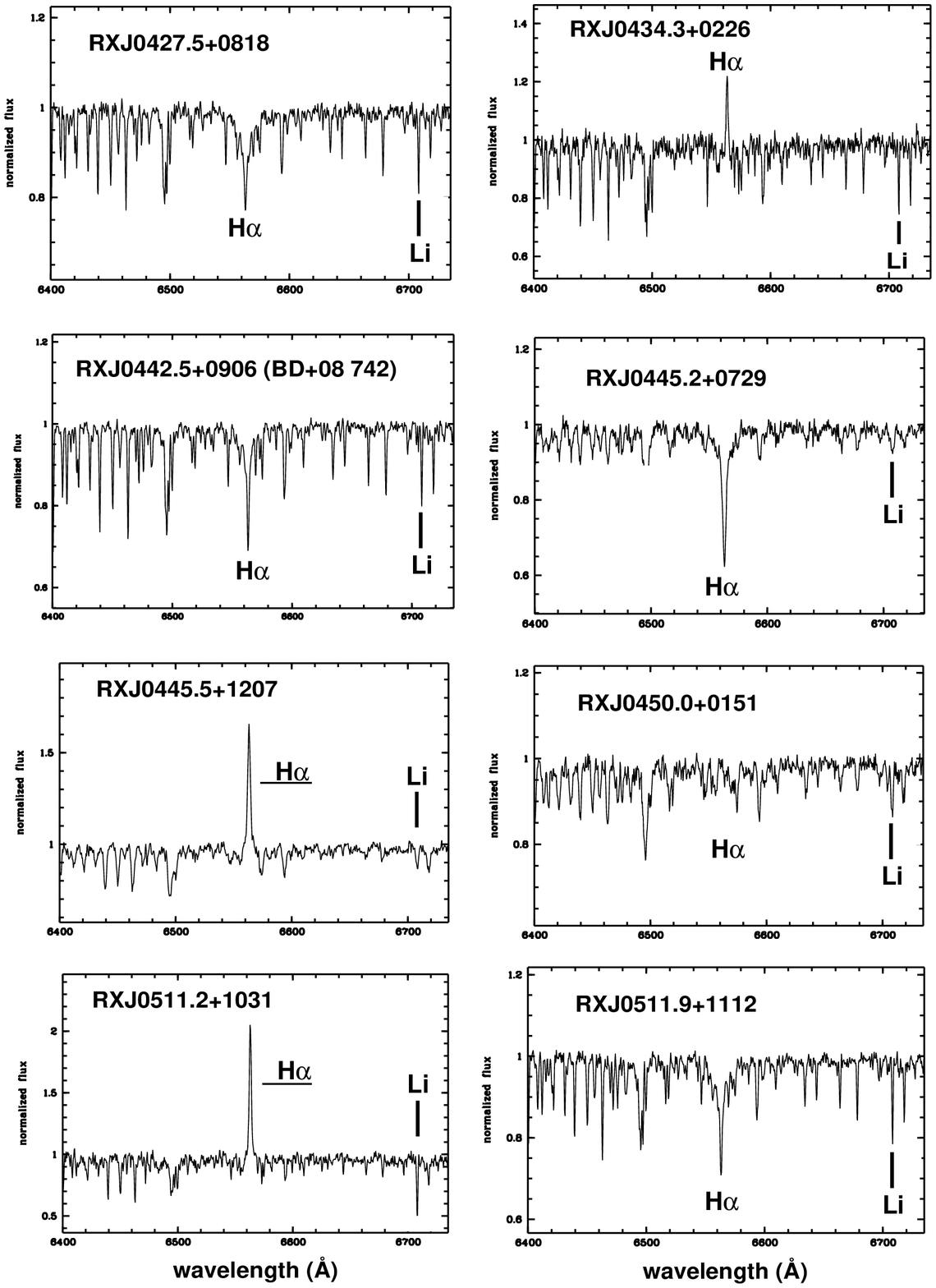,height=22.5 cm}
\caption[]{Continued}
 \end{figure*}

\addtocounter{figure}{-1}

\begin{figure*}
\centerline{\psfig {file=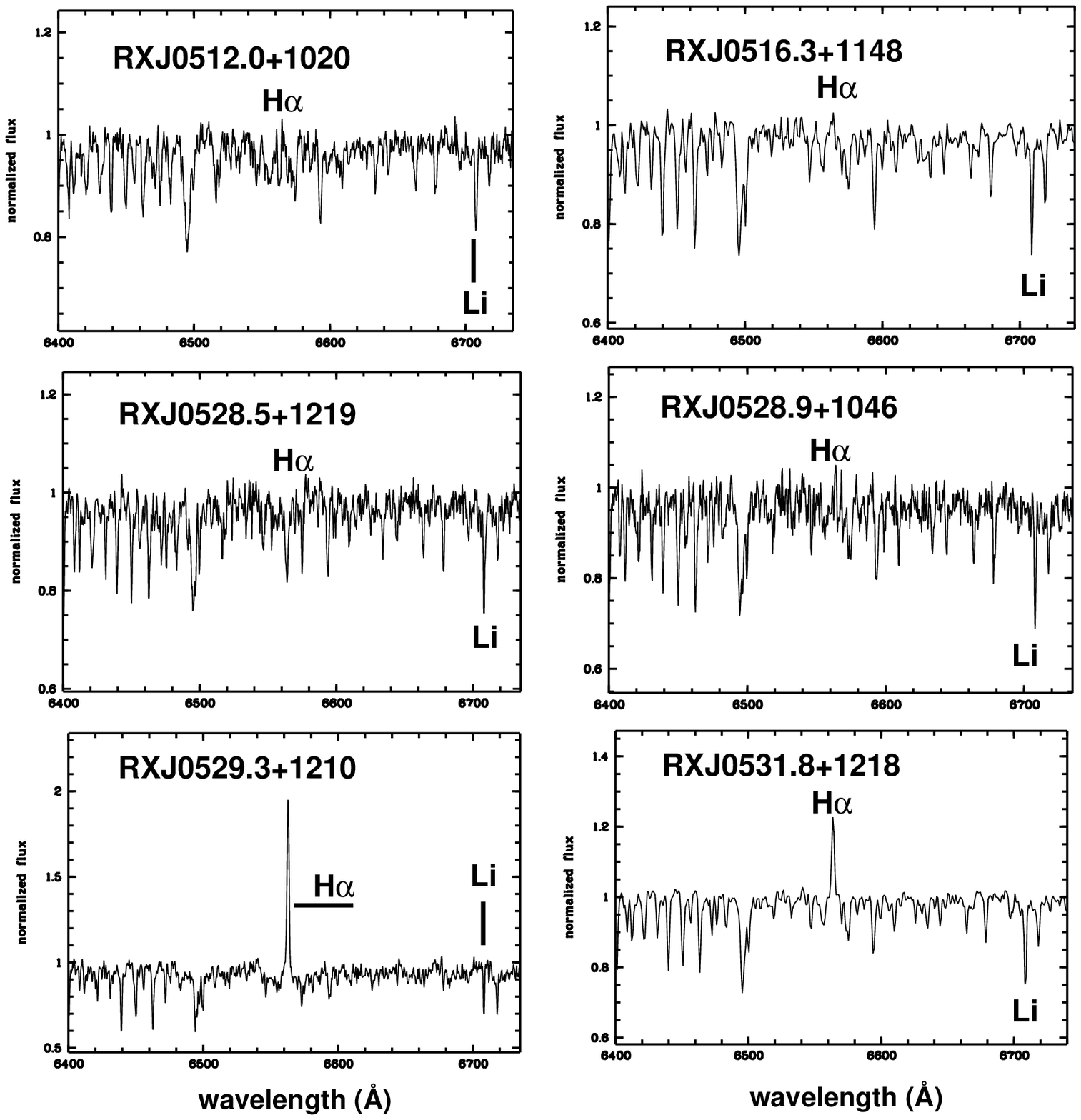,bbllx=40pt,bblly=142pt,bburx=561pt,bbury=680pt,clip=,height=18.2 cm}}
\caption[]{Continued}
 \end{figure*}

Most of the stars classified here as PMS? and some of those labelled as
PMS are probably not very young (as indicated by their relatively
shallow Li line) and may be close to the Main Sequence.  Hence, they
may well be the long sought post-TTS which have been dispersed out of
the region of ongoing star formation.  If star formation has been going
on in Taurus-Auriga for $\sim 10^{7}$ years, there should be numerous
post-TTS all around the CO clouds, even if the velocity dispersion in
Taurus-Auriga is only 1-2~km~s$^{-1}$ (Jones \& Herbig 1979). On the
other hand, considering the strength of the \li\  doublet as a youth
indicator, there seem to be very young new wTTS in our sample.  If they
are less than $10^{6}$ years old, it appears to be impossible for them
to have moved the distance between central Taurus-Auriga and their
present location by slow isotropic drifting. If they have been formed
in central Taurus-Auriga, they must have moved to their present
location with high velocities.

Two mechanisms have been proposed to explain the existence of  very
young objects so far away from the SFRs.  According to the first one,
these objects, called ``run-away'' TTS (RATTS: Sterzik \etl\ 1995,
Neuh\"auser \etl\ 1995c), can be ejected by three-body encounters in
multiple protostellar systems (Sterzik \& Durisen 1995).  Many  RATTS
are expected in the vicinity of other SFRs as the pre-selection of TTS
candidates indicates (e.g. Sterzik \etl\ 1995). A different explanation
has been proposed by Feigelson (1996), who argues that TTS can form in
small, high-velocity, short-lived cloudlets within and around a
turbulent giant molecular cloud complex (like Taurus-Auriga and
Chamaeleon).

Our sample of new low-mass PMS stars found south of the Taurus-Auriga
dark cloud complex may contain relatively old dispersed post-TTS, young
wTTS formed locally in small short-lived cloudlets, and young wTTS
ejected from the central areas of ongoing star formation.  With
kinematic data (both radial velocities and proper motions) and age
estimates (e.g. from placing the stars into the HR-diagram and from
precise Li abundances) one may be able to distinguish between these
different contributions in the future. The presence of lithium in a
large number of stars in the general direction of any known SFR studied
is an observational fact that certainly needs further explanation.

\begin {acknowledgements} We would like to thank Guillermo Torres for
fruitful discussions.  Many thanks also to the EXSAS and ROSAT teams at
MPE as well as to the support teams at the different telescope sites
used in this research.  This work has made use of the SIMBAD database
operated at CDS, Strasbourg.  The ROSAT project has been supported by
the Bundesministerium f\"ur Bildung, Wissenschaft, Forschung und
Technologie (BMBW/DARA) and the Max-Planck-Gesellschaft. 
\end {acknowledgements}

\end{document}